\documentclass[aps,prl,groupedaddress,twocolumn,showpacs,amsmath,amssymb,floatfix]{revtex4-1}

\usepackage{xspace,amsmath,amsfonts,amsthm,amssymb,amsbsy,amstext}
\usepackage{graphicx}
\usepackage{amssymb}
\usepackage{color}
\usepackage{psfrag}
\usepackage{ifsym}
\usepackage{epstopdf}
\usepackage{bbold}
\usepackage{mathrsfs}
\usepackage[usenames,dvipsnames]{xcolor}
\usepackage{latexsym}
\usepackage[colorlinks=true,citecolor=Blue,linkcolor=Purple, urlcolor=Cerulean]{hyperref}
\usepackage{bbm}
\DeclareMathAlphabet\mathbfcal{OMS}{cmsy}{b}{n}

\definecolor{pink}{rgb}{1,0.078,0.57}

\definecolor{green}{rgb}{0,0.7,0.9}

\newcommand{\ket}[1] {\left\vert #1 \right\rangle}

\newcommand{\br}{\mathbf{r}}
\newcommand{\bE}{\mathbf{E}}
\newcommand{\be}{\hat{\mathbf{e}}}
\newcommand{\dg}{^{\dagger}}
\newcommand{\Tr}{\mathrm{Tr}}
\newcommand{\hc}{\mathrm{h.c.}}

\newcommand{\vac}{\ket{\mathrm{vac}}}

\newcommand{\uv}[1]{\hat{\mathbf{#1}}}

\newcommand{\mI}{\mathbf{I}}
\newcommand{\mE}{\mathbf{E}}
\newcommand{\mJ}{\mathbf{J}}

\newcommand{\mV}{\mathbf{V}}

\begin{document}

\title{Construction of Multichromophoric Spectra from Monomer Data:\\
 Applications to Resonant Energy Transfer}

\author{Aur\'elia Chenu}
\email{achenu@mit.edu}
\author{Jianshu Cao}
\email{jianshu@mit.edu}
\affiliation{Massachusetts Institute of Technology, 77 Massachusetts Avenue, Cambridge, MA 02139, USA}


 \begin{abstract} 
We develop a model that establishes a quantitative link between the physical properties of molecular aggregates and their constituent building blocks. The relation is built on the coherent potential approximation, calibrated against exact results, and proven reliable for a wide range of  parameters. It provides a practical method to compute spectra and transfer rates in multichromophoric systems from experimentally accessible monomer data. Applications to F\"orster energy transfer reveal optimal transfer rates as functions of both the system-bath coupling and intra-aggregate coherence. 

\end{abstract}

\maketitle
Molecular self-assembly is used in nature to build complex structures and regulate material properties \cite{Douglas2009a,*Mayer2002a, *Kos2009a}, and can also be  exploited for the fabrication of versatile nanostructures \cite{Whitesides1991a,*Zhang2003a}. 
The spatial arrangement of chromophoric building blocks strongly influences the electronic distribution \cite{Scherf2002a}, enabling a broad range of optical and transport properties \cite{ValkunasBook}. 
Nature has mastered this art, evolving from a very limited number of monomers an impressive diversity of photosynthetic light-harvesting complexes \cite{Cogdell2008a}, 
which are known to be highly versatile \cite{Hu1998a, *Scholes2012a} and efficient in absorbing sunlight and transferring the subsequent excitation \cite{ValkunasBook,McConnell2010a}.
Understanding the sensitive interplay between monomer and super-structure (composed of monomers), and its influence on the optical, electronic and transport properties is highly desirable
for the synthesis of new materials \cite{Scherf2002a}, the design and operation of organic-based devices \cite{Spano2014a}, including solar cells \cite{Bredas2009a, *Heeger2010a, *Goubard2015a}, transistors, light-emitting diodes \cite{SherfBook}, and flexible electronics \cite{Orgiu2014a}. 
Yet despite its fundamental role, the relationship between molecular super-structure and physical properties lacks systematic quantitative understanding. 
In this Letter, we derive such a quantitative method, by establishing the relation between the aggregate spectra and its constituent monomer building blocks;
we further calibrate it against exact results, and apply the theory to the important dynamical process of resonant energy transfer.

Optical excitations of organic compounds involve both electronic and vibrational degrees of freedom \cite{Levinson1973a}.
While the exciton-phonon interaction is well understood for monomers \cite{MukamelBook,MahanBook,Broude1985a}, the electronic coupling in super-structures such as multichromophoric (MC) complexes delocalizes the excitation \cite{Davydov1962a} and therefore requires the treatment of electron-vibrational coupling, excitonic coupling and disorder on an equal footing \cite{Chenu2015a}. 
The available techniques, either exact such as stochastic path integral (sPI) \cite{Moix2015a} and hierarchical equation of motions (HEOM) \cite{Ishizaki2009c,Struempfer2011a,Jing2013a}, or approximate such as full-cumulant-expansion (FCE) \cite{Ma2015a}, 2nd-order time convolution \cite{Renger2000a}, time convolutionless \cite{Renger2002a, Banchi2013a} and other recent developments \cite{Dinh2015a,Gelzinis2015a}, are computationally expensive and not universally applicable to relate the structure to optical properties. 
  These treatments require microscopic Hamiltonians and thus are not explicit about the structure--spectra relation. Our approach establishes such a relation: it allows us to predict the physical properties of complex structures or, conversely, infer the structure from its measured properties.

While construction of the optical properties is important in its own right, the spectra also provide additional transport information. The transfer rates between weakly coupled excitonic systems can be obtained from the overlap of the donor emission and acceptor absorption spectra using F\"orster resonant energy transfer (FRET) \cite{Forster1965a}. The original FRET theory describes the environment through its effect on the monomer spectra. Extensions to MC systems \cite{Sumi1999a, Mukai1999a, Scholes2003a,Jang2004a}, where the donor/acceptor are composed of coupled chromophores, demonstrated that the far-field linear spectroscopic line shapes are insufficient; rather, the near-field, polarization-resolved aggregate spectra are needed to obtain the MCFT (MC Fluorescent Transfer) rate in general. 
Though nonlinear spectroscopic experiments \cite{Scholes2000a} could, in principle, be used, the required information is not accessible with current experimental techniques \cite{Jang2004a}. 
The theory developed here solves this problem, allowing for the construction of the aggregate spectra from the experimentally accessible monomer spectra. 

Our model is based on the coherent potential approximation (CPA) \cite{Sumi1974a}: it treats the vibrational coupling exactly at the monomer level and includes all orders of electronic coupling, treated exactly up to the second order and approximately for higher orders. 
Benchmarks against exact sPI \cite{Moix2015a} and FCE  \cite{Ma2015a}  calculations show that our model is reliable over a wide range of parameters.
Our theory applies to MCFT and recovers some aspects of the classical treatments \cite{Kuhn1970a, Chance1975a, Zimanyi2010a, Duque2015a}  as a limiting case. 
It completes the series of papers quantifying the reliability of different quantum models in MC systems \cite{Ma2015a, Ma2015b, Moix2015a}.

\emph{Monomer spectra.---}We introduce the notation and exact formulation of monomer spectra using the independent boson model \cite{MahanBook}.  
Consider a single monomer coupled to a thermal phonon bath, both labeled by $n$, and characterized by the spin-boson Hamiltonian 
\begin{equation}\label{eq:H0_n}
H_0^{(n)} = H^{(n)}_{0S} + H^{(n)}_{B} + H^{(n)}_{SB} ,
\end{equation}
where $H^{(n)}_{0S}= E_n B\dg_n B_n$ denotes the Hamiltonian of the electronic system, $H_B^{(n)} = \sum_k \hbar \omega_{n,k} b\dg_{n,k} b_{n,k}$ that of the bath, and  $H^{(n)}_{SB}= B\dg_n B_n \sum_k g_{n,k} (b\dg_{n,k} + b_{n,k})$ their coupling, which is taken linear in the bath coordinate.
The operator $B\dg_n$ creates an excitation on monomer $n$, forming the state $B\dg_n \vac = \ket{n}$; $b\dg_{n,k}$ creates a phononic excitation in mode $k$. 
The excited state energy $E_n = \hbar \omega^0_n + \lambda_n$ includes the reorganization energy $\lambda_n =\sum_k g_{n,k} / \omega_{n,k}$.
Interaction with the electric field is treated semi-classically through the system-radiation interaction Hamiltonian $H_{SR}^{(n)} = \hat{\mu}_n \cdot E\be$, where $\hat{\mu}_n = \vec{\mu}_{ng} B\dg_n  + \hc$ denotes the transition dipole moment operator and $\be$ is a unit vector. 

Following Sumi \cite{Sumi1974a}, we use the retarded Green's function  \cite{MahanBook}
  associated with the monomer Hamiltonian (\ref{eq:H0_n}),
 \begin{equation}\label{eq:G0n}
\mathcal{G}^0_{nn}(t) = -\frac{i}{\hbar} \Theta(t) e^{\frac{i}{\hbar} H_0^{(n)} t} B_n\, e^{-\frac{i}{\hbar} H_0^{(n)}t} B\dg_n,
\end{equation}
to define the optical spectra---all given in units of energy here. 
The absorption spectrum is obtained from the imaginary part of the Green's function averaged over the phonon bath,  $\left\langle \mathcal{G}^0_{nn}(\omega) \right\rangle_g$, where $\langle \bullet \rangle_{g} \equiv \Tr_B[ \bullet \rho^0_{g}] $
denotes the trace over the bath using the density matrix of the system-bath in its ground state, $\rho^{0}_g$. 
The experimentally accessible spectrum includes the dipole transition,
$I^{(n)}_{\textrm{exp}} (\omega)= (\be \cdot \vec{\mu}_{gn}) \, I_0^{(n)}(\omega)  \left( \vec{\mu}_{ng} \cdot \be \right)$, 
where  
\begin{equation} \label{eq:abs_mon}
I_0^{(n)}(\omega) = -2  {\rm Im} \int_{-\infty}^{\infty} dt \: e^{i \omega t} \,  \Tr_B \left[ \mathcal{G}^0_{nn}(t)\rho^{0}_g \right].
\end{equation}

This monomer spectrum can be evaluated exactly as follows. Assuming a Franck-Condon transition from the ground state, the initial state can be taken as the factorized state $\rho^0_g = \mathbbm{1}_S \otimes \rho_B$, where  $\rho_B = e^{- \beta H_B^{(n)}}  / \Tr[e^{-\beta H_B^{(n)}}]$ is the bath density matrix at equilibrium, with  $\beta^{-1} = k_B T$. For a harmonic bath at thermal equilibrium, the absorption lineshape (\ref{eq:abs_mon}) is exactly
\begin{equation} \label{eq:abs_exact}
 I_0^{(n)}(\omega)  = \frac{2}{\hbar} {\rm Re} \int_{0}^{\infty} dt \, e^{i\omega t} e^{-i \omega_{ng} t} e^{-g_n(t) },
 \end{equation}
where $\omega_{ng}  \equiv (E_n-E_g) / \hbar$ and $E_g$ denotes the electronic ground state energy. The lineshape function $g_n(t) = \int_0^t d\tau_1 \int_0^{\tau_2} d\tau_2 C_n(\tau_2) $ is obtained from the bath auto-correlation function 
$C_n(\tau) = \frac{1}{\hbar^2} \langle H_{SB}^{(n)}(\tau) H_{SB}^{(n)}(0) \rangle$, and can be evaluated exactly assuming a Drude spectral density, $C_n''(\omega) \equiv 2 \lambda_n \Lambda \omega / (\omega^2 + \Lambda^2)$,  with  $\Lambda$ the cutoff frequency \cite{MukamelBook}.

Steady-state emission occurs after the entire system-bath has equilibrated within the single-exciton manifold and is obtained from $\left\langle \mathcal{G}^0_{nn}(\omega) \right\rangle_e$, with $\langle \bullet \rangle_{e} \equiv \Tr_B[ \bullet \rho^0_{e}] $. 
The initial state $\rho^0_e$ entering the averaged Green's function is not factorized anymore, and the system-bath entanglement needs to be considered \cite{Ma2015a}. 
Instead of a direct calculation, 
we follow Refs. \cite{Sumi1999a, Banchi2013a, Moix2015a} and use the detailed balance condition  \cite{VanKampenBook} to obtain the emission spectrum from the absorption,
\begin{equation} \label{eq:emi_DB}
E_0^{(n)}(\omega) = \frac{e^{\beta \hbar \omega}}{Z^{(n)}}  I_0^{(n)}(\omega),
\end{equation}
where $Z^{(n)} = \Tr [e^{- \beta H_{0S}^{(n)}} e^{- \beta H_{B}^{(n)}}]$ is the monomer partition function. 

\emph{Multichromophoric spectra.---}Consider now a system of $N$ coupled chromophores described by 
\begin{equation}\label{eq:Htot}
H = H_0 + V,
\end{equation}
where $H_0 = \sum_{n=1}^N H_0^{(n)}$ is the sum over $N$ independent monomers (\ref{eq:H0_n}), 
which includes exciton-phonon interaction,
and $V = \sum_{n\neq m} V_{nm} B\dg_n B_m$ characterizes the inter-monomer coupling, typically of dipole-dipole nature. 
The operator $B\dg_n$
now denotes excitation of the $n$th monomer exclusively. 
Interaction with light is now characterized by $H_{SR} = \sum_n H_{SR}^{(n)}$. 
We denote $\mathbfcal{G}$ and $\mathbfcal{G}^0$ the $N\times N$ matrices formed by the  Green's functions associated, respectively, with the total Hamiltonian $H$ and the unperturbed Hamiltonian $H_0$, which matrix elements are  
\begin{subequations}
\begin{align}
\mathcal{G}_{nn'}(t) &=  -\frac{i}{\hbar} \Theta(t) e^{\frac{i}{\hbar} H t } B_n\, e^{-\frac{i}{\hbar} H t} B\dg_{n'} \label{eq:G_def}\\
\mathcal{G}^0_{nn'}(t) &=  -\frac{i}{\hbar} \Theta(t) e^{\frac{i}{\hbar} H_0 t } B_n\, e^{-\frac{i}{\hbar} H_0 t} B\dg_{n'}  \label{eq:G0_def}.
\end{align}
\end{subequations}
Note that the diagonal elements of $\mathbfcal{G}^0$ are equal to the monomer Green's functions (\ref{eq:G0n}). 
In principle, the MC Green's function $\mathbfcal{G}$ can be exactly expressed using the unperturbed Green's function $\mathbfcal{G}^0$ and the self energy according to Dysons' equation \cite{MahanBook}. 
Tracing over the phonon bath would then provide the MC absorption and emission tensors,
$\mI(\omega)= -2 \: {\rm Im}  \left\langle \mathbfcal{G}(\omega) \right \rangle_g $ and  $\mE(\omega)= -2 \: {\rm Im}  \left\langle \mathbfcal{G}(\omega) \right \rangle_e $, respectively. 
However, while an exact solution exists for single monomers (\ref{eq:H0_n}),
the inter-monomer coupling $V$ in MC systems (\ref{eq:Htot}) tends to delocalize the electronic excitation and mix the vibrational and electronic degrees of freedom. Evaluating the trace then requires methods numerically expensive  \cite{Moix2015a}, and often approximate \cite{Ma2015a, Ma2015b}.

The approximate theory derived here provides an analytical expression in terms of the constituent spectra and structural properties, allowing for an explicit relation between the optical properties and the structure. Using the integral representation of the exponential, the bath-averaged total Green's function can be expanded, in the time or frequency domain, as \cite{CPA_SI2016}
\begin{equation}\label{eq:G_expansion}
\langle \mathbfcal{G} \rangle = \langle \mathbfcal{G}^0 \rangle + \langle \mathbfcal{G}^0 V \mathbfcal{G}^0 \rangle + \langle \mathbfcal{G}^0 V \mathbfcal{G}^0 V \mathbfcal{G}^0  \rangle + \dots,
\end{equation}
 where the $\langle \bullet \rangle$ denote the trace over the bath and the subscript ($g$ or $e$) characterizing the initial state will be specified as needed.
The 0th order term is simply the diagonal matrix of the monomeric Green's function (\ref{eq:G0n}) with the proper initial state, i.e., $\langle \mathbfcal{G}^0 \rangle_{g/e} =  (\langle \mathcal{G}^0_{nn} \rangle_{g/e} )$. 
The first-order term can be evaluated exactly for the factorized initial state, $\langle \bullet \rangle_g$,  because (i) the trace then commutes with the bath density operator, (ii) there is one and only one electronic transition involved, and (iii) the individual baths are uncorrelated.
The trace can thus be split exactly,
\begin{equation}\label{eq:1order}
\langle \mathbfcal{G}^0 V \mathbfcal{G}^0 \rangle_g=  \langle \mathbfcal{G}^0 \rangle_g \mV  \langle \mathbfcal{G}^0 \rangle_g ,
\end{equation}
where $\mV=(V_{nm})$ is a tensor. 
For the second-order term in Eq. (\ref{eq:G_expansion}), we neglect the phonon correlations and take
\begin{equation} \label{eq:approx}
 \left \langle \mathcal{G}^0_{nn}(\omega) \mathcal{G}^0_{nn}(\omega) \right \rangle_g \approx \left \langle \mathcal{G}^0_{nn}(\omega) \right \rangle_g \left \langle \mathcal{G}^0_{nn}(\omega) \right \rangle_g.
 \end{equation}
This approximation is analogous to the decoupling scheme used in the single-site dynamical coherent potential approximation (CPA) \cite{Broude1985a} and will be referred as such.
 It is the main approximation here, yielding to our key result. It allows simplifying all higher orders such that the full Green's function (\ref{eq:G_expansion}) with the initial ground-state density matrix reduces to: 
\begin{equation}\label{eq:G_CPA}
\langle \mathbfcal{G}(\omega) \rangle_g \approx \frac{\left \langle \mathbfcal{G}^0(\omega) \right \rangle_g }{\mathbbm{1}_N -  \mV \left \langle \mathbfcal{G}^0(\omega) \right \rangle_g}.
\end{equation}
 The CPA approach treats the bath coupling exactly at the monomer level. It is exact up to the second-order of intra-aggregate coupling $V$ and includes all higher orders approximately. As such, our approach is more robust than other methods derived for weak coupling \cite{Yang2005a, Mancal2011a}, especially in highly delocalized cases.

The MC absorption tensor is then simply 
\begin{equation} \label{eq:I_CPA}
\mI_{\rm CPA}(\omega)= -2 \: {\rm Im}  \frac{\left \langle \mathbfcal{G}^0(\omega) \right \rangle_g }{\mathbbm{1}_N -  \mV \left \langle \mathbfcal{G}^0(\omega) \right \rangle_g}.
\end{equation}
The far-field, measurable absorption spectrum, is given by 
$I_\textrm{exp}(\omega) =  \sum_{n n'}  (\uv{e} \cdot \vec{\mu}_n )  \:  I_{nn'}(\omega) \: ( \vec{\mu}_{n'} \cdot \uv{e})$.
Note that, in calculating the full MC tensor, the CPA requires knowledge of both the real and imaginary parts of the monomeric Green's function $ \left \langle \mathbfcal{G}^0(\omega) \right \rangle_g$. The latter is accessible experimentally through the absorption spectra of the constituent monomers and their transition dipole moments; the real part is related through the Kramer-Kronig relation \cite{MukamelBook}. The CPA approach therefore allows constructing the MC absorption tensor  (\ref{eq:I_CPA}) from its monomer features, either experimentally or theoretically accessible.
%---we remind that $\Tr_B[\mathbfcal{G}^0 (\omega) \rho^0_g]$ has a closed-form solution assuming a thermal bath with a Drude spectral density, \textit{cf} (\ref{eq:abs_exact}). 
Also, we show in Ref. \cite{CPA_SI2016} that Eq. (\ref{eq:I_CPA}) reduces to the tensor derived from a classical picture of oscillating dipoles \cite{Zimanyi2010a, Duque2015a}, when the coupling is restricted to dipole-dipole interaction. This suggests that the CPA approximation (\ref{eq:approx}) is implicit in the classical electrostatic treatment of absorption \cite{Cao1993a}.

\begin{figure}
\hspace*{-.05\columnwidth}
\includegraphics[width=1.1\columnwidth]{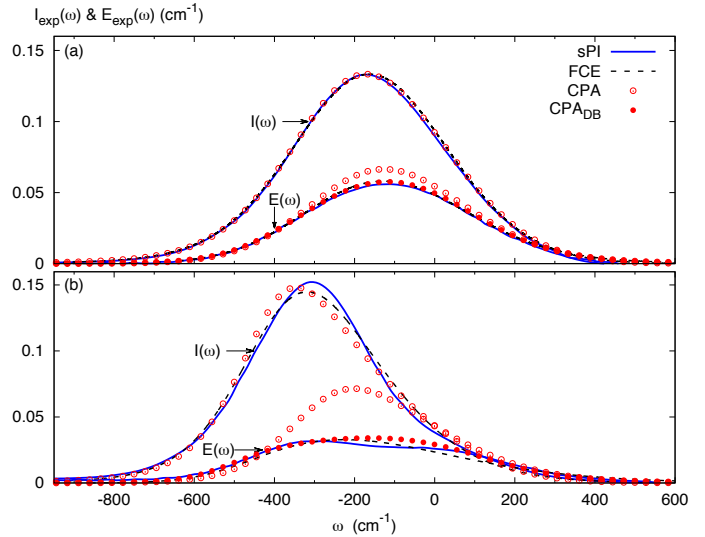}\caption{Comparison of models for MC absorption $I_{\rm exp}(\omega)$ and emission $E_{\rm exp}(\omega)$ spectra, for (a) localized and (b) delocalized dimers. Absorption  (\ref{eq:I_CPA})  obtained from the CPA treatment developed here well matches sPI exact   \cite{Moix2015a} and  FCE approximate \cite{Ma2015a} results; emission spectra (\ref{eq:E_CPA}) resemble the sum of the individual spectra (not shown here), and are not accurate for large inter-chromophore couplings (b). Adding detailed balance, the CPA$_{\rm DB}$ (\ref{eq:E_CPADB}) provides accurate results over a wide range of couplings. The difference between CPA and CPA$_{\rm DB}$ displays the influence of the system-bath entanglement, which is important for emission. 
Parameters correspond to (a) case I (V = 20 cm$^{-1}$, $\Delta E_{21}\equiv E_2 - E_1$= 100 cm$^{-1}$) and (b) case II (V = 100 cm$^{-1}$, $\Delta E_{21}$ = 20 cm$^{-1}$) in Ref. \cite{Ma2015a} with $\lambda$=100 cm$^{-1}$,  $\Lambda$=53 cm$^{-1}$, $T$=300 K, and $\uv{e}$ along $\vec{\mu}$. \label{fig:spectra} }
\end{figure}

Direct application of the CPA (\ref{eq:approx}) with ($g\rightarrow e$) yields the emission tensor 
\begin{equation}\label{eq:E_CPA}
\mE_{\rm CPA}(\omega)= -2 \: {\rm Im}  \frac{\left \langle \mathbfcal{G}^0(\omega) \right \rangle_e }{\mathbbm{1}_N +  \mV \left \langle \mathbfcal{G}^0(\omega) \right \rangle_e},
\end{equation}
where the initial density matrix is the equilibrium state in the first-excited manifold $\rho^0_e$. 
Because of the initial system-bath entanglement, the separation of averaging (\ref{eq:1order}) with $(g\rightarrow e)$  is no longer exact, and the CPA is approximate already in the first order of $V$ for the emission tensor. Numerical simulations show that the prediction is similar to the sum of the monomer emission spectra (\ref{eq:emi_DB}), and therefore deviates from the exact solution for strong coupling [Fig. \ref{fig:spectra}(b)]. 
Instead of Eq. (\ref{eq:E_CPA}) and similarly to the monomer treatment (\ref{eq:emi_DB}), we calculate the emission tensor from the detailed balance (DB), which applies to the total system as
\begin{equation} \label{eq:E_CPADB}
\mE(\omega) = \frac{e^{\beta \hbar\omega }}{\Tr[e^{- \beta H}]} \: \mI(\omega).
\end{equation}
We label this emission tensor by ``CPA$_{\rm DB}$'' when using Eq. (\ref{eq:I_CPA}) for the absorption tensor $\mI(\omega)$.  
 The normalization factor can be obtained either from direct sPI calculation \cite{Moix2012a} or from the absorption spectrum 
 using the mirror property of the spectra, i.e., $\mE(\omega^0 - \lambda - \omega) = \mI(\omega^0 + \lambda + \omega ) / Z_{0S}$, where $Z_{0S} \equiv \Tr(e^{-\beta \sum_n H_{0S}^{(n)}})$ is given by the  monomer  system Hamiltonian  \cite{MukamelBook, Moix2015a}. 
The emission spectrum is then $E_\textrm{exp}(\omega) =  \uv{e} \cdot (\vec{\mu}^T . \bE(\omega) . \vec{\mu}) \cdot \uv{e}$.

Figure \ref{fig:spectra} presents the absorption and emission spectra for the two dimers detailed in Ref. \cite{Ma2015a}, i.e. for weak and strong inter-chromophore coupling $V$. It is shown that the proposed treatment [Eqs. (\ref{eq:I_CPA}-\ref{eq:E_CPADB})] provides accurate predictions for both spectra, even for relatively strong coupling---$V/\lambda =1$ in Fig. \ref{fig:spectra}(b). The CPA with detailed balance (\ref{eq:E_CPADB}) greatly enhances the results over the CPA only (\ref{eq:E_CPA}), thereby showing the importance of the bath's first-order correlation function when the initial state is the system-bath entangled density matrix.
Comparisons with the FCE over a wider range of parameters are presented in Ref. \cite{CPA_SI2016}.

\emph{Application to energy transfer rate.---}Knowledge of the spectral tensors allows for the determination of the transfer rate between a donor ($D$) and an acceptor  ($A$) aggregate using Fermi's golden rule. We consider a system of $M$-coupled donor and $N$-coupled acceptor chromophores described by the total Hamiltonian
\begin{equation} \label{eq:H_AD}
H^{AD}  = H^A + H^D + J^{AD},
\end{equation}
where the MC Hamiltonian of the donor $H^D = H_0^D + V^D$  and that of the acceptor $H^A = H_0^A + V^A$ is described by (\ref{eq:Htot}), changing $B_n \to D_n (A_n)$, respectively, and where the inter-chromophore coupling is $V^D= \sum_{m\neq m'}^M V^D_{m m'} D\dg_m D_{m'}$ and  $V^A = \sum_{n\neq n'}^N V^A_{n n'} A\dg_n A_{n'}$. 
$J^{AD}$ denotes the coupling between the donor-acceptor chromophores, i.e., $J^{AD} = \sum_{n}^N \sum_{m}^M J^{AD}_{nm} A\dg_n D_m + \hc$ The operators $D\dg_m$ and $A\dg_n$, respectively, denote excitation of the donor monomer $m$ and the acceptor monomer $n$.

The rate of multichromophoric F\"orster resonant energy transfer is given by the overlap of the emission and absorption tensors \cite{Ma2015a, Banchi2013a, Jang2004a},
\begin{equation} \label{eq:k_def}
k = \int_{-\infty}^{\infty}  \frac{d\omega}{2\pi} \: \Tr\left[ \mJ^T \mE^D(\omega) \mJ \: \mI^A(\omega)\right],
\end{equation}
where the matrix $\mJ = (J^{AD}_{nm})$ denotes the donor-acceptor coupling strength. The emission and absorption tensors, respectively $\mE^D(\omega) = [E^{D}_{mm'}(\omega)]$ and $\mI^A(\omega) = [I^{A}_{nn'}(\omega)]$, are the polarization-resolved near-field spectral components, which are known to be necessary in MC systems for significant intra-donor ($V^D$) or intra-acceptor ($V^A$) couplings  \cite{Jang2004a}. 

\begin{figure}
\hspace*{-0.05\columnwidth}
\includegraphics[width=1.05\columnwidth]{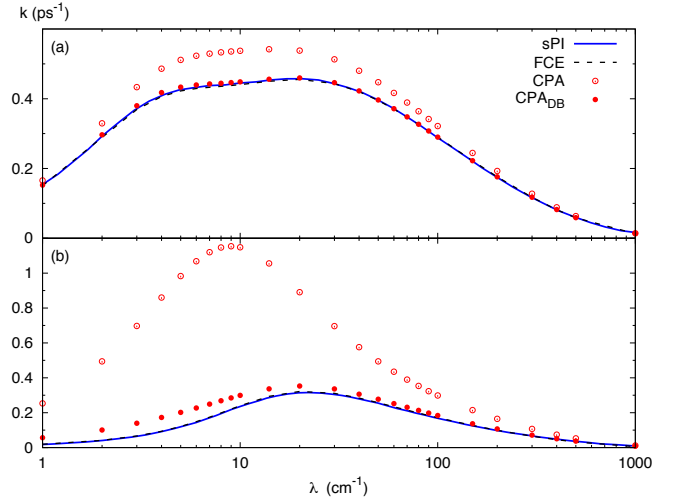}
\caption{Energy transfer rate (\ref{eq:rate_CPA}) for different bath reorganization energies $\lambda$  using the developed CPA and CPA$_{\rm DB}$ models to calculate absorption (\ref{eq:I_CPA}) and emission (\ref{eq:E_CPA}, \ref{eq:E_CPADB}) tensors for localized (a) and delocalized (b) systems. Comparison with exact sPI results \cite{Moix2015a} show perfect matching of the CPA$_{\rm DB}$ for small electronic coupling (a), and a slight over-prediction of the rate for large coupling (b). The error using only CPA comes from overpredicting the emission tensor (cf. Fig. \ref{fig:spectra}). We used $J^{AD}_{nm} = 10$ cm$^{-1}$; $\lambda^D = \lambda^A$ and $V^A_{nn'} = V^D_{mm'} = V$. $\Delta E_{21}^A = \Delta E_{21}^D = \Delta E_{21}$ are as in Fig. (\ref{fig:spectra}).  \label{fig:rate}}
\end{figure}

Using the derived treatment, specifically the CPA absorption tensor (\ref{eq:I_CPA}) for the acceptor  along with the CPA$_{\rm DB}$ emission tensor (\ref{eq:E_CPADB}) for the donor, the MCFT rate becomes  
\begin{eqnarray} \label{eq:rate_CPA}
k &\approx& \int_{-\infty}^{\infty} \frac{d\omega}{2\pi} \: \Tr\left[ \mJ^T  \frac{e^{\beta \hbar\omega }}{\Tr[e^{- \beta H^D}]}   2{\rm Im}  \left(  \frac{\langle\mathbfcal{G}^0_D(\omega) \rangle_g }{\mathbbm{1}_M -  \mV^D \: \langle \mathbfcal{G}^0_D(\omega) \rangle_g }\right) \right.  \nonumber  \\
&&\left. \times \mJ \:  2{\rm Im} \left( \frac{\langle \mathbfcal{G}^0_A(\omega)\rangle_g }{\mathbbm{1}_N -  \mV^A \: \langle \mathbfcal{G}^0_A(\omega) \rangle_g } \right) \right], 
\end{eqnarray}
where $\mathbfcal{G}^0_D$ ($\mathbfcal{G}^0_A$) is a $M\times M$ ($N\times N$) matrix formed by the Green's functions of the uncoupled monomers constituting the donor (acceptor) aggregate, i.e., defined by Eq. (\ref{eq:G0_def}) changing $B_n \to D_n (A_n)$. 
This rate expression only requires the monomer bath-averaged Green's functions $\langle \mathbfcal{G}^0 \rangle_g$, which includes the system-bath coupling exactly at the monomer level and can be evaluated exactly for a thermal bath (\ref{eq:abs_exact}) or determined experimentally. All influence from electronic coupling is contained in the matrices describing intra-donor $ \mV^D$, intra-acceptor $ \mV^A$ and inter donor-acceptor $ \mJ^{AD}$ couplings, and not restricted to dipole-dipole coupling. The rate (\ref{eq:rate_CPA}) is exact up to second order in the intra-aggregate couplings $\mV$ and includes all higher orders approximately.

Figure (\ref{fig:rate}) presents the transfer rate for localized and delocalized donor/acceptor (cases I\&II in Ref. \cite{Ma2015a}, respectively) for different reorganization energies $\lambda$. Comparison with the exact path-integral calculations shows perfect agreement for the localized case (\ref{fig:rate}a), and a slight overprediction  for highly delocalized MC systems Fig. (\ref{fig:rate}(b).

\begin{figure}
\includegraphics[width=\columnwidth]{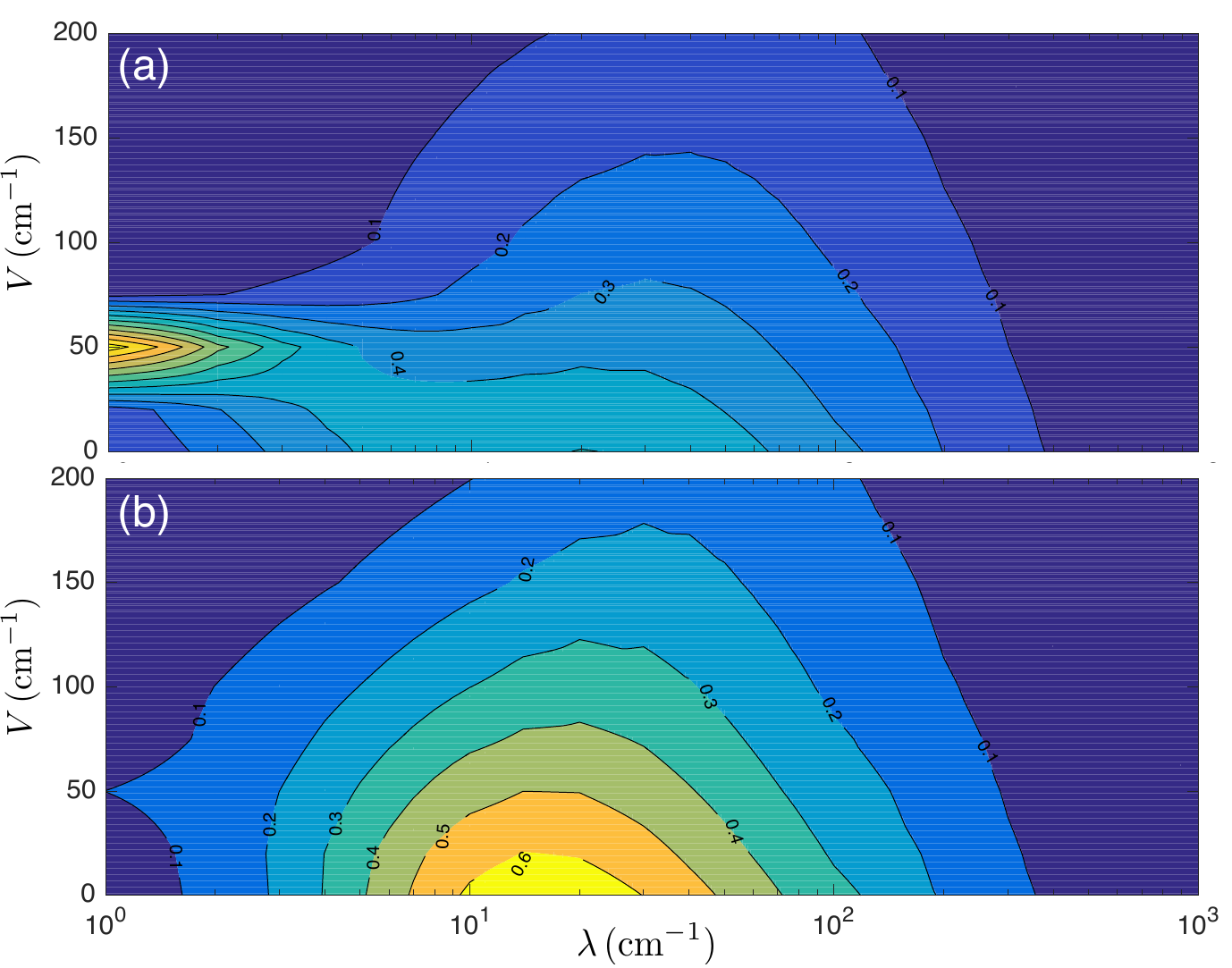}
\caption{ Contour map of the transfer rate as function of the reorganization energy $\lambda = \lambda^A = \lambda^D$ and the intra-aggregate coupling strength $V = V^A_{nn'} = V^D_{mm'}$. Panel (a) exhibits an optimal intra-aggregate coupling strength around $V\sim 50$cm$^{-1}$, and panel (b) clearly shows the optimal bath coupling.  The different behaviors between (a) and (b) steam from different electronic splitting $\Delta E_{12}^{A(D)}$, respectively, (a) 100 and (b)  20 cm$^{-1}$. \label{fig:rate_map}}
\end{figure}

The simplicity of our approach [Eq. (\ref{eq:rate_CPA})] allows predicting the transfer rates over a wide range of structural parameters. Figure  (\ref{fig:rate_map}) shows the rate as a function of the reorganization energy and intra-aggregate coupling $V$ for systems with different electronic splittings $\Delta E_{21}$.
%, respectively   20 cm$^{-1}$ and  100 cm$^{-1}$. 
We clearly see an optimal bath-coupling strength, confirming environment-assisted quantum transport  \cite{Chorosajev2014a}. 
Interestingly, Fig. \ref{fig:rate_map}(a) also exhibits an optimal intra-aggregate coherence (V$\sim$50 cm$^{-1}$), which was not previously reported. 
The existence of such optimum depends on the system configuration, as seen from the comparison with rates in a system with smaller energy gap, Fig. \ref{fig:rate_map}(b). While this dependance requires further investigation, our results confirm that intra-aggregate couplings can enhance transfer, which is in line with Refs. \cite{Jang2004a,Cheng2006a, Zimanyi2010a, Kim2010b,Wu2013b}. 

 In summary, we extended the applicability of the CPA to absorption and emission tensors of multichromophoric systems, and showed accurate results over a surprisingly wide range of structure parameters. This approach now allows for a reliable prediction of the MCFT rate, which reveals that, additionally to optimal environment couplings, the intra-aggregate coupling can be optimized to enhance transport.
Our treatment  identifies the correction terms, and recovers the classical absorption tensor as a limiting case, suggesting that first-order bath correlations are neglected classically. 
Our model could be further extended to include the off-diagonal bath coupling, introduced through electronic coupling, using, e.g., the two-particle dynamical CPA \cite{Friesner1982a, *Lagos1984a}.

Beyond fast and reliable characterization of multi-chromophic complexes, a quantitative relation between physical properties and aggregate structure is established. 
 This straightforward approach is based on spectroscopic measurements and does not require a microscopic Hamiltonian. It allows us to explore a large space of structure parameters and optimize the aggregate structure based on its optical and transport properties. As such, we anticipate that it will be a relevant tool to  experimentally and theoretically  describe electronic excitation and excitonic energy transfer. 
\begin{acknowledgments}  
\emph{Acknowledgments.---}We thank P. Brumer for interesting discussions on the topic and A. del Campo for comments on the manuscript. We acknowledge funding from the Swiss National Science Foundation (A.C.) and the NSF (Grant No. CHE-1112825).
\end{acknowledgments}

%\bibliography{main} 

\begin{thebibliography}{53}%
\makeatletter
\providecommand \@ifxundefined [1]{%
 \@ifx{#1\undefined}
}%
\providecommand \@ifnum [1]{%
 \ifnum #1\expandafter \@firstoftwo
 \else \expandafter \@secondoftwo
 \fi
}%
\providecommand \@ifx [1]{%
 \ifx #1\expandafter \@firstoftwo
 \else \expandafter \@secondoftwo
 \fi
}%
\providecommand \natexlab [1]{#1}%
\providecommand \enquote  [1]{``#1''}%
\providecommand \bibnamefont  [1]{#1}%
\providecommand \bibfnamefont [1]{#1}%
\providecommand \citenamefont [1]{#1}%
\providecommand \href@noop [0]{\@secondoftwo}%
\providecommand \href [0]{\begingroup \@sanitize@url \@href}%
\providecommand \@href[1]{\@@startlink{#1}\@@href}%
\providecommand \@@href[1]{\endgroup#1\@@endlink}%
\providecommand \@sanitize@url [0]{\catcode `\\12\catcode `\$12\catcode
  `\&12\catcode `\#12\catcode `\^12\catcode `\_12\catcode `\%12\relax}%
\providecommand \@@startlink[1]{}%
\providecommand \@@endlink[0]{}%
\providecommand \url  [0]{\begingroup\@sanitize@url \@url }%
\providecommand \@url [1]{\endgroup\@href {#1}{\urlprefix }}%
\providecommand \urlprefix  [0]{URL }%
\providecommand \Eprint [0]{\href }%
\providecommand \doibase [0]{http://dx.doi.org/}%
\providecommand \selectlanguage [0]{\@gobble}%
\providecommand \bibinfo  [0]{\@secondoftwo}%
\providecommand \bibfield  [0]{\@secondoftwo}%
\providecommand \translation [1]{[#1]}%
\providecommand \BibitemOpen [0]{}%
\providecommand \bibitemStop [0]{}%
\providecommand \bibitemNoStop [0]{.\EOS\space}%
\providecommand \EOS [0]{\spacefactor3000\relax}%
\providecommand \BibitemShut  [1]{\csname bibitem#1\endcsname}%
\let\auto@bib@innerbib\@empty
%</preamble>
\bibitem [{\citenamefont {Douglas}\ \emph {et~al.}(2009)\citenamefont
  {Douglas}, \citenamefont {Dietz},\ and\ \citenamefont
  {Liedl}}]{Douglas2009a}%
  \BibitemOpen
  \bibfield  {author} {\bibinfo {author} {\bibfnamefont {S.~M.}\ \bibnamefont
  {Douglas}}, \bibinfo {author} {\bibfnamefont {H.}~\bibnamefont {Dietz}}, \
  and\ \bibinfo {author} {\bibfnamefont {T.}~\bibnamefont {Liedl}},\ }\href
  {http://dx.doi.org/10.1038/nature08016} {\bibfield  {journal} {\bibinfo
  {journal} {Nature}\ }\textbf {\bibinfo {volume} {459}},\ \bibinfo {pages}
  {414} (\bibinfo {year} {2009})}\BibitemShut {NoStop}%
\bibitem [{\citenamefont {Mayer}\ and\ \citenamefont
  {Sarikaya}(2002)}]{Mayer2002a}%
  \BibitemOpen
  \bibfield  {author} {\bibinfo {author} {\bibfnamefont {G.}~\bibnamefont
  {Mayer}}\ and\ \bibinfo {author} {\bibfnamefont {M.}~\bibnamefont
  {Sarikaya}},\ }\href {\doibase 10.1007/BF02412144} {\bibfield  {journal}
  {\bibinfo  {journal} {Exp. Mech.}\ }\textbf {\bibinfo {volume} {42}},\
  \bibinfo {pages} {395} (\bibinfo {year} {2002})}\BibitemShut {NoStop}%
\bibitem [{\citenamefont {Kos}\ and\ \citenamefont {Ford}(2009)}]{Kos2009a}%
  \BibitemOpen
  \bibfield  {author} {\bibinfo {author} {\bibfnamefont {V.}~\bibnamefont
  {Kos}}\ and\ \bibinfo {author} {\bibfnamefont {R.}~\bibnamefont {Ford}},\
  }\href {\doibase 10.1007/s00018-009-0064-9} {\bibfield  {journal} {\bibinfo
  {journal} {Cellular and Molecular Life Sciences}\ }\textbf {\bibinfo {volume}
  {66}},\ \bibinfo {pages} {311} (\bibinfo {year} {2009})}\BibitemShut
  {NoStop}%
\bibitem [{\citenamefont {Whitesides}\ \emph {et~al.}(1991)\citenamefont
  {Whitesides}, \citenamefont {Mathias},\ and\ \citenamefont
  {Seto}}]{Whitesides1991a}%
  \BibitemOpen
  \bibfield  {author} {\bibinfo {author} {\bibfnamefont {G.}~\bibnamefont
  {Whitesides}}, \bibinfo {author} {\bibfnamefont {J.}~\bibnamefont {Mathias}},
  \ and\ \bibinfo {author} {\bibfnamefont {C.}~\bibnamefont {Seto}},\ }\href
  {\doibase 10.1126/science.1962191} {\bibfield  {journal} {\bibinfo  {journal}
  {Science}\ }\textbf {\bibinfo {volume} {254}},\ \bibinfo {pages} {1312}
  (\bibinfo {year} {1991})}\BibitemShut {NoStop}%
\bibitem [{\citenamefont {Zhang}(2003)}]{Zhang2003a}%
  \BibitemOpen
  \bibfield  {author} {\bibinfo {author} {\bibfnamefont {S.}~\bibnamefont
  {Zhang}},\ }\href {http://dx.doi.org/10.1038/nbt874} {\bibfield  {journal}
  {\bibinfo  {journal} {Nat. Biotech.}\ }\textbf {\bibinfo {volume} {21}},\
  \bibinfo {pages} {1171} (\bibinfo {year} {2003})}\BibitemShut {NoStop}%
\bibitem [{\citenamefont {Scherf}\ and\ \citenamefont
  {List}(2002)}]{Scherf2002a}%
  \BibitemOpen
  \bibfield  {author} {\bibinfo {author} {\bibfnamefont {U.}~\bibnamefont
  {Scherf}}\ and\ \bibinfo {author} {\bibfnamefont {E.}~\bibnamefont {List}},\
  }\href {\doibase 10.1002/1521-4095(20020404)14:7<477::AID-ADMA477>3.0.CO;2-9}
  {\bibfield  {journal} {\bibinfo  {journal} {Adv. Mat.}\ }\textbf {\bibinfo
  {volume} {14}},\ \bibinfo {pages} {477} (\bibinfo {year} {2002})}\BibitemShut
  {NoStop}%
\bibitem [{\citenamefont {van Amerongen}\ \emph {et~al.}(2000)\citenamefont
  {van Amerongen}, \citenamefont {Valkunas},\ and\ \citenamefont {van
  Grondelle}}]{ValkunasBook}%
  \BibitemOpen
  \bibfield  {author} {\bibinfo {author} {\bibfnamefont {H.}~\bibnamefont {van
  Amerongen}}, \bibinfo {author} {\bibfnamefont {L.}~\bibnamefont {Valkunas}},
  \ and\ \bibinfo {author} {\bibfnamefont {R.}~\bibnamefont {van Grondelle}},\
  }\href@noop {} {\emph {\bibinfo {title} {Photosynthetic Excitons}}}\
  (\bibinfo  {publisher} {World Scientific},\ \bibinfo {address} {Singapore},\
  \bibinfo {year} {2000})\BibitemShut {NoStop}%
\bibitem [{\citenamefont {Cogdell}\ \emph {et~al.}(2008)\citenamefont
  {Cogdell}, \citenamefont {Gardiner}, \citenamefont {Hashimoto},\ and\
  \citenamefont {Brotosudarmo}}]{Cogdell2008a}%
  \BibitemOpen
  \bibfield  {author} {\bibinfo {author} {\bibfnamefont {R.~J.}\ \bibnamefont
  {Cogdell}}, \bibinfo {author} {\bibfnamefont {A.~T.}\ \bibnamefont
  {Gardiner}}, \bibinfo {author} {\bibfnamefont {H.}~\bibnamefont {Hashimoto}},
  \ and\ \bibinfo {author} {\bibfnamefont {T.~H.~P.}\ \bibnamefont
  {Brotosudarmo}},\ }\href {\doibase 10.1039/B807201A} {\bibfield  {journal}
  {\bibinfo  {journal} {Photochem. Photobiol. Sci.}\ }\textbf {\bibinfo
  {volume} {7}},\ \bibinfo {pages} {1150} (\bibinfo {year} {2008})}\BibitemShut
  {NoStop}%
\bibitem [{\citenamefont {Hu}\ \emph {et~al.}(1998)\citenamefont {Hu},
  \citenamefont {Damjanovi\'c}, \citenamefont {Ritz},\ and\ \citenamefont
  {Schulten}}]{Hu1998a}%
  \BibitemOpen
  \bibfield  {author} {\bibinfo {author} {\bibfnamefont {X.}~\bibnamefont
  {Hu}}, \bibinfo {author} {\bibfnamefont {A.}~\bibnamefont {Damjanovi\'c}},
  \bibinfo {author} {\bibfnamefont {T.}~\bibnamefont {Ritz}}, \ and\ \bibinfo
  {author} {\bibfnamefont {K.}~\bibnamefont {Schulten}},\ }\href {\doibase
  10.1073/pnas.95.11.5935} {\bibfield  {journal} {\bibinfo  {journal} {Proc.
  Natl. Acad. Sci. U. S. A.}\ }\textbf {\bibinfo {volume} {95}},\ \bibinfo
  {pages} {5935} (\bibinfo {year} {1998})}\BibitemShut {NoStop}%
\bibitem [{\citenamefont {Scholes}\ \emph {et~al.}(2012)\citenamefont
  {Scholes}, \citenamefont {Mirkovic}, \citenamefont {Turner}, \citenamefont
  {Fassioli},\ and\ \citenamefont {Buchleitner}}]{Scholes2012a}%
  \BibitemOpen
  \bibfield  {author} {\bibinfo {author} {\bibfnamefont {G.~D.}\ \bibnamefont
  {Scholes}}, \bibinfo {author} {\bibfnamefont {T.}~\bibnamefont {Mirkovic}},
  \bibinfo {author} {\bibfnamefont {D.~B.}\ \bibnamefont {Turner}}, \bibinfo
  {author} {\bibfnamefont {F.}~\bibnamefont {Fassioli}}, \ and\ \bibinfo
  {author} {\bibfnamefont {A.}~\bibnamefont {Buchleitner}},\ }\href {\doibase
  10.1039/C2EE23013E} {\bibfield  {journal} {\bibinfo  {journal} {Energy
  Environ. Sci.}\ }\textbf {\bibinfo {volume} {5}},\ \bibinfo {pages} {9374}
  (\bibinfo {year} {2012})}\BibitemShut {NoStop}%
\bibitem [{\citenamefont {McConnell}\ \emph {et~al.}(2010)\citenamefont
  {McConnell}, \citenamefont {Li},\ and\ \citenamefont
  {Brudvig}}]{McConnell2010a}%
  \BibitemOpen
  \bibfield  {author} {\bibinfo {author} {\bibfnamefont {I.}~\bibnamefont
  {McConnell}}, \bibinfo {author} {\bibfnamefont {G.}~\bibnamefont {Li}}, \
  and\ \bibinfo {author} {\bibfnamefont {G.~W.}\ \bibnamefont {Brudvig}},\
  }\href {\doibase http://dx.doi.org/10.1016/j.chembiol.2010.05.005} {\bibfield
   {journal} {\bibinfo  {journal} {Chem. \& Bio.}\ }\textbf {\bibinfo {volume}
  {17}},\ \bibinfo {pages} {434} (\bibinfo {year} {2010})}\BibitemShut
  {NoStop}%
\bibitem [{\citenamefont {Spano}\ and\ \citenamefont
  {Rivas}(2014)}]{Spano2014a}%
  \BibitemOpen
  \bibfield  {author} {\bibinfo {author} {\bibfnamefont {F.}~\bibnamefont
  {Spano}}\ and\ \bibinfo {author} {\bibfnamefont {C.}~\bibnamefont {Rivas}},\
  }\href {\doibase 10.1146/annurev-physchem-040513-103639} {\bibfield 
   {journal} {\bibinfo  {journal} {Annu. Rev. Phys.
  Chem.}\ }\textbf {\bibinfo {volume} {65}},\ \bibinfo {pages} {477} (\bibinfo
  {year} {2014})}\BibitemShut {NoStop}%
\bibitem [{\citenamefont {Br\'edas}\ \emph {et~al.}(2009)\citenamefont
  {Br\'edas}, \citenamefont {Norton}, \citenamefont {Cornil},\ and\
  \citenamefont {Coropceanu}}]{Bredas2009a}%
  \BibitemOpen
  \bibfield  {author} {\bibinfo {author} {\bibfnamefont {J.-L.}\ \bibnamefont
  {Br\'edas}}, \bibinfo {author} {\bibfnamefont {J.~E.}\ \bibnamefont
  {Norton}}, \bibinfo {author} {\bibfnamefont {J.}~\bibnamefont {Cornil}}, \
  and\ \bibinfo {author} {\bibfnamefont {V.}~\bibnamefont {Coropceanu}},\
  }\href {\doibase 10.1021/ar900099h} {\bibfield  {journal} {\bibinfo
  {journal} {Acc. Chem. Res.}\ }\textbf {\bibinfo {volume} {42}},\ \bibinfo
  {pages} {1691} (\bibinfo {year} {2009})}\BibitemShut {NoStop}%
\bibitem [{\citenamefont {Heeger}(2010)}]{Heeger2010a}%
  \BibitemOpen
  \bibfield  {author} {\bibinfo {author} {\bibfnamefont {A.}~\bibnamefont
  {Heeger}},\ }\href {\doibase 10.1039/B914956M} {\bibfield  {journal}
  {\bibinfo  {journal} {Chem. Soc. Rev.}\ }\textbf {\bibinfo {volume} {39}},\
  \bibinfo {pages} {2354} (\bibinfo {year} {2010})}\BibitemShut {NoStop}%
\bibitem [{\citenamefont {Goubard}\ and\ \citenamefont
  {Dumur}(2015)}]{Goubard2015a}%
  \BibitemOpen
  \bibfield  {author} {\bibinfo {author} {\bibfnamefont {F.}~\bibnamefont
  {Goubard}}\ and\ \bibinfo {author} {\bibfnamefont {F.}~\bibnamefont
  {Dumur}},\ }\href {\doibase 10.1039/C4RA11559G} {\bibfield  {journal}
  {\bibinfo  {journal} {RSC Adv.}\ }\textbf {\bibinfo {volume} {5}},\ \bibinfo
  {pages} {3521} (\bibinfo {year} {2015})}\BibitemShut {NoStop}%
\bibitem [{\citenamefont {Sherf}(2006)}]{SherfBook}%
  \BibitemOpen
  \bibfield  {author} {\bibinfo {author} {\bibfnamefont {K.~M.~U.}\
  \bibnamefont {Sherf}},\ }\href@noop {} {\emph {\bibinfo {title} {Organic
  {L}ight {E}mitting {D}evices: {S}ynthesis, {P}roperties and
  {A}pplications}}}\ (\bibinfo  {publisher} {New York: Wiley},\ \bibinfo {year}
  {2006})\BibitemShut {NoStop}%
\bibitem [{\citenamefont {Orgiu}\ and\ \citenamefont
  {Samor\'{i}}(2014)}]{Orgiu2014a}%
  \BibitemOpen
  \bibfield  {author} {\bibinfo {author} {\bibfnamefont {E.}~\bibnamefont
  {Orgiu}}\ and\ \bibinfo {author} {\bibfnamefont {P.}~\bibnamefont
  {Samor\'{i}}},\ }\href {\doibase 10.1002/adma.201304695} {\bibfield
  {journal} {\bibinfo  {journal} {Adv. Mat.}\ }\textbf {\bibinfo {volume}
  {26}},\ \bibinfo {pages} {1827} (\bibinfo {year} {2014})}\BibitemShut
  {NoStop}%
\bibitem [{\citenamefont {Levinson}\ and\ \citenamefont
  {Rashba}(1973)}]{Levinson1973a}%
  \BibitemOpen
  \bibfield  {author} {\bibinfo {author} {\bibfnamefont {Y.}~\bibnamefont
  {Levinson}}\ and\ \bibinfo {author} {\bibfnamefont {E.}~\bibnamefont
  {Rashba}},\ }\href {http://iopscience.iop.org/0034-4885/36/12/001} {\bibfield
   {journal} {\bibinfo  {journal} {Rep. Prog. Phys.}\ }\textbf {\bibinfo
  {volume} {36}},\ \bibinfo {pages} {1499} (\bibinfo {year}
  {1973})}\BibitemShut {NoStop}%
\bibitem [{\citenamefont {Mukamel}(1995)}]{MukamelBook}%
  \BibitemOpen
  \bibfield  {author} {\bibinfo {author} {\bibfnamefont {S.}~\bibnamefont
  {Mukamel}},\ }\href@noop {} {\emph {\bibinfo {title} {Principles of nonlinear
  spectroscopy}}}\ (\bibinfo  {publisher} {Oxford University Press},\ \bibinfo
  {address} {Oxford},\ \bibinfo {year} {1995})\ \bibinfo {note} {{C}h.
  8}\BibitemShut {NoStop}%
\bibitem [{\citenamefont {Mahan}(2000)}]{MahanBook}%
  \BibitemOpen
  \bibfield  {author} {\bibinfo {author} {\bibfnamefont {G.~D.}\ \bibnamefont
  {Mahan}},\ }\href@noop {} {\emph {\bibinfo {title} {Many-{P}article
  {P}hysics}}},\ \bibinfo {edition} {3rd}\ ed.\ (\bibinfo  {publisher} {Kluwer
  Academic Publishers},\ \bibinfo {year} {2000})\ \bibinfo {note} {{C}h. 2 and Ch.
  4}\BibitemShut {NoStop}%
\bibitem [{\citenamefont {Broude}\ \emph {et~al.}(1985)\citenamefont {Broude},
  \citenamefont {Rashba},\ and\ \citenamefont {Sheka}}]{Broude1985a}%
  \BibitemOpen
  \bibfield  {author} {\bibinfo {author} {\bibfnamefont {V.}~\bibnamefont
  {Broude}}, \bibinfo {author} {\bibfnamefont {E.}~\bibnamefont {Rashba}}, \
  and\ \bibinfo {author} {\bibfnamefont {E.}~\bibnamefont {Sheka}},\
  }\href@noop {} {\emph {\bibinfo {title} {Spectroscopy of Molecular
  Excitons}}},\ edited by\ \bibinfo {editor} {\bibfnamefont {V.}~\bibnamefont
  {Goldanskii}}\ (\bibinfo  {publisher} {Springer Series in Chemical Physics},\
  \bibinfo {year} {1985})\BibitemShut {NoStop}%
\bibitem [{\citenamefont {Davydov}(1962)}]{Davydov1962a}%
  \BibitemOpen
  \bibfield  {author} {\bibinfo {author} {\bibfnamefont {A.~S.}\ \bibnamefont
  {Davydov}},\ }\href@noop {} {\emph {\bibinfo {title} {Theory of molecular
  excitons}}}\ (\bibinfo  {publisher} {McGraw-Hill, New York},\ \bibinfo {year}
  {1962})\BibitemShut {NoStop}%
\bibitem [{\citenamefont {Chenu}\ and\ \citenamefont
  {Scholes}(2015)}]{Chenu2015a}%
  \BibitemOpen
  \bibfield  {author} {\bibinfo {author} {\bibfnamefont {A.}~\bibnamefont
  {Chenu}}\ and\ \bibinfo {author} {\bibfnamefont {G.}~\bibnamefont
  {Scholes}},\ }\href {\doibase 10.1146/annurev-physchem-040214-121713}
  {\bibfield  {journal} {\bibinfo  {journal} {Annu. Rev. Phys. Chem.}\ }\textbf
  {\bibinfo {volume} {66}},\ \bibinfo {pages} {69} (\bibinfo {year}
  {2015})}\BibitemShut {NoStop}%
\bibitem [{\citenamefont {Moix}\ \emph {et~al.}(2015)\citenamefont {Moix},
  \citenamefont {Ma},\ and\ \citenamefont {Cao}}]{Moix2015a}%
  \BibitemOpen
  \bibfield  {author} {\bibinfo {author} {\bibfnamefont {J.~M.}\ \bibnamefont
  {Moix}}, \bibinfo {author} {\bibfnamefont {J.}~\bibnamefont {Ma}}, \ and\
  \bibinfo {author} {\bibfnamefont {J.}~\bibnamefont {Cao}},\ }\href {\doibase
  10.1063/1.4908601} {\bibfield  {journal} {\bibinfo  {journal} {J. Chem.
  Phys.}\ }\textbf {\bibinfo {volume} {142}},\ \bibinfo {pages} {094108}
  (\bibinfo {year} {2015})}\BibitemShut {NoStop}%
\bibitem [{\citenamefont {Ishizaki}\ and\ \citenamefont
  {Fleming}(2009)}]{Ishizaki2009c}%
  \BibitemOpen
  \bibfield  {author} {\bibinfo {author} {\bibfnamefont {A.}~\bibnamefont
  {Ishizaki}}\ and\ \bibinfo {author} {\bibfnamefont {G.~R.}\ \bibnamefont
  {Fleming}},\ }\href {\doibase 10.1063/1.3155372} {\bibfield  {journal}
  {\bibinfo  {journal} {J. Chem. Phys.}\ }\textbf {\bibinfo {volume} {130}},\
  \bibinfo {pages} {234111} (\bibinfo {year} {2009})}\BibitemShut {NoStop}%
\bibitem [{\citenamefont {Str\"{u}mpfer}\ and\ \citenamefont
  {Schulten}(2011)}]{Struempfer2011a}%
  \BibitemOpen
  \bibfield  {author} {\bibinfo {author} {\bibfnamefont {J.}~\bibnamefont
  {Str\"{u}mpfer}}\ and\ \bibinfo {author} {\bibfnamefont {K.}~\bibnamefont
  {Schulten}},\ }\href {\doibase 10.1063/1.3557042} {\bibfield  {journal}
  {\bibinfo  {journal} {J. Chem. Phys.}\ }\textbf {\bibinfo {volume} {134}},\
  \bibinfo {pages} {095102} (\bibinfo {year} {2011})}\BibitemShut {NoStop}%
\bibitem [{\citenamefont {Jing}\ \emph {et~al.}(2013)\citenamefont {Jing},
  \citenamefont {Chen}, \citenamefont {Bai},\ and\ \citenamefont
  {Shi}}]{Jing2013a}%
  \BibitemOpen
  \bibfield  {author} {\bibinfo {author} {\bibfnamefont {Y.}~\bibnamefont
  {Jing}}, \bibinfo {author} {\bibfnamefont {L.}~\bibnamefont {Chen}}, \bibinfo
  {author} {\bibfnamefont {S.}~\bibnamefont {Bai}}, \ and\ \bibinfo {author}
  {\bibfnamefont {Q.}~\bibnamefont {Shi}},\ }\href {\doibase
  http://dx.doi.org/10.1063/1.4775843} {\bibfield  {journal} {\bibinfo
  {journal} {J. Chem. Phys.}\ }\textbf {\bibinfo {volume} {138}},\ \bibinfo
  {eid} {045101} (\bibinfo {year} {2013})}\BibitemShut {NoStop}%
\bibitem [{\citenamefont {Ma}\ and\ \citenamefont {Cao}(2015)}]{Ma2015a}%
  \BibitemOpen
  \bibfield  {author} {\bibinfo {author} {\bibfnamefont {J.}~\bibnamefont
  {Ma}}\ and\ \bibinfo {author} {\bibfnamefont {J.}~\bibnamefont {Cao}},\
  }\href {\doibase 10.1063/1.4908599} {\bibfield  {journal} {\bibinfo
  {journal} {J. Chem. Phys.}\ }\textbf {\bibinfo {volume} {142}},\ \bibinfo
  {pages} {094106} (\bibinfo {year} {2015})}\BibitemShut {NoStop}%
\bibitem [{\citenamefont {Renger}\ and\ \citenamefont
  {May}(2000)}]{Renger2000a}%
  \BibitemOpen
  \bibfield  {author} {\bibinfo {author} {\bibfnamefont {T.}~\bibnamefont
  {Renger}}\ and\ \bibinfo {author} {\bibfnamefont {V.}~\bibnamefont {May}},\
  }\href {\doibase 10.1103/PhysRevLett.84.5228} {\bibfield  {journal} {\bibinfo
   {journal} {Phys. Rev. Lett.}\ }\textbf {\bibinfo {volume} {84}},\ \bibinfo
  {pages} {5228} (\bibinfo {year} {2000})}\BibitemShut {NoStop}%
\bibitem [{\citenamefont {Renger}\ and\ \citenamefont
  {Marcus}(2002)}]{Renger2002a}%
  \BibitemOpen
  \bibfield  {author} {\bibinfo {author} {\bibfnamefont {T.}~\bibnamefont
  {Renger}}\ and\ \bibinfo {author} {\bibfnamefont {R.~A.}\ \bibnamefont
  {Marcus}},\ }\href {\doibase 10.1063/1.1470200} {\bibfield  {journal}
  {\bibinfo  {journal} {J. Chem. Phys.}\ }\textbf {\bibinfo {volume} {116}},\
  \bibinfo {pages} {9997} (\bibinfo {year} {2002})}\BibitemShut {NoStop}%
\bibitem [{\citenamefont {Banchi}\ \emph {et~al.}(2013)\citenamefont {Banchi},
  \citenamefont {Costagliola}, \citenamefont {Ishizaki},\ and\ \citenamefont
  {Giorda}}]{Banchi2013a}%
  \BibitemOpen
  \bibfield  {author} {\bibinfo {author} {\bibfnamefont {L.}~\bibnamefont
  {Banchi}}, \bibinfo {author} {\bibfnamefont {G.}~\bibnamefont {Costagliola}},
  \bibinfo {author} {\bibfnamefont {A.}~\bibnamefont {Ishizaki}}, \ and\
  \bibinfo {author} {\bibfnamefont {P.}~\bibnamefont {Giorda}},\ }\href
  {\doibase 10.1063/1.4803694} {\bibfield  {journal} {\bibinfo  {journal} {J.
  Chem. Phys.}\ }\textbf {\bibinfo {volume} {138}},\ \bibinfo {pages} {184107}
  (\bibinfo {year} {2013})}\BibitemShut {NoStop}%
\bibitem [{\citenamefont {Dinh}\ and\ \citenamefont
  {Renger}(2015)}]{Dinh2015a}%
  \BibitemOpen
  \bibfield  {author} {\bibinfo {author} {\bibfnamefont {T.-C.}\ \bibnamefont
  {Dinh}}\ and\ \bibinfo {author} {\bibfnamefont {T.}~\bibnamefont {Renger}},\
  }\href {\doibase http://dx.doi.org/10.1063/1.4904928} {\bibfield  {journal}
  {\bibinfo  {journal} {J. Chem. Phys.}\ }\textbf {\bibinfo {volume} {142}},\
  \bibinfo {eid} {034104} (\bibinfo {year} {2015})}\BibitemShut {NoStop}%
\bibitem [{\citenamefont {Gelzinis}\ \emph {et~al.}(2015)\citenamefont
  {Gelzinis}, \citenamefont {Abramavicius},\ and\ \citenamefont
  {Valkunas}}]{Gelzinis2015a}%
  \BibitemOpen
  \bibfield  {author} {\bibinfo {author} {\bibfnamefont {A.}~\bibnamefont
  {Gelzinis}}, \bibinfo {author} {\bibfnamefont {D.}~\bibnamefont
  {Abramavicius}}, \ and\ \bibinfo {author} {\bibfnamefont {L.}~\bibnamefont
  {Valkunas}},\ }\href {\doibase http://dx.doi.org/10.1063/1.4918343}
  {\bibfield  {journal} {\bibinfo  {journal} {J. Chem. Phys.}\ }\textbf
  {\bibinfo {volume} {142}},\ \bibinfo {eid} {154107} (\bibinfo {year}
  {2015})}\BibitemShut {NoStop}%
\bibitem [{\citenamefont {F\"orster}(1965)}]{Forster1965a}%
  \BibitemOpen
  \bibfield  {author} {\bibinfo {author} {\bibfnamefont {T.}~\bibnamefont
  {F\"orster}},\ }in\ \href@noop {} {\emph {\bibinfo {booktitle} {Moden Quantum
  Chemistry}}},\ Vol.\ \bibinfo {volume} {{III}.},\ \bibinfo {editor} {edited
  by\ \bibinfo {editor} {\bibfnamefont {O.}~\bibnamefont {Sinano\v{g}lu}}}\
  (\bibinfo {year} {1965})\ p.~\bibinfo {pages} {93}\BibitemShut {NoStop}%
\bibitem [{\citenamefont {Sumi}(1999)}]{Sumi1999a}%
  \BibitemOpen
  \bibfield  {author} {\bibinfo {author} {\bibfnamefont {H.}~\bibnamefont
  {Sumi}},\ }\href {\doibase 10.1021/jp983477u} {\bibfield  {journal} {\bibinfo
   {journal} {J. Phys. Chem. B}\ }\textbf {\bibinfo {volume} {103}},\ \bibinfo
  {pages} {252} (\bibinfo {year} {1999})}\BibitemShut {NoStop}%
\bibitem [{\citenamefont {Mukai}\ \emph {et~al.}(1999)\citenamefont {Mukai},
  \citenamefont {Abe}, ,\ and\ \citenamefont {Sumi}}]{Mukai1999a}%
  \BibitemOpen
  \bibfield  {author} {\bibinfo {author} {\bibfnamefont {K.}~\bibnamefont
  {Mukai}}, \bibinfo {author} {\bibfnamefont {S.}~\bibnamefont {Abe}}, , \ and\
  \bibinfo {author} {\bibfnamefont {H.}~\bibnamefont {Sumi}},\ }\href {\doibase
  10.1021/jp984469g} {\bibfield  {journal} {\bibinfo  {journal} {J. Phys. Chem.
  B}\ }\textbf {\bibinfo {volume} {103}},\ \bibinfo {pages} {6096} (\bibinfo
  {year} {1999})}\BibitemShut {NoStop}%
\bibitem [{\citenamefont {Scholes}(2003)}]{Scholes2003a}%
  \BibitemOpen
  \bibfield  {author} {\bibinfo {author} {\bibfnamefont {G.~D.}\ \bibnamefont
  {Scholes}},\ }\href {\doibase 10.1146/annurev.physchem.54.011002.103746}
  {\bibfield  {journal} {\bibinfo  {journal} {Annu. Rev. Phys. Chem.}\ }\textbf
  {\bibinfo {volume} {54}},\ \bibinfo {pages} {57} (\bibinfo {year}
  {2003})}\BibitemShut {NoStop}%
\bibitem [{\citenamefont {Jang}\ \emph {et~al.}(2004)\citenamefont {Jang},
  \citenamefont {Newton},\ and\ \citenamefont {Silbey}}]{Jang2004a}%
  \BibitemOpen
  \bibfield  {author} {\bibinfo {author} {\bibfnamefont {S.}~\bibnamefont
  {Jang}}, \bibinfo {author} {\bibfnamefont {M.~D.}\ \bibnamefont {Newton}}, \
  and\ \bibinfo {author} {\bibfnamefont {R.~J.}\ \bibnamefont {Silbey}},\
  }\href {\doibase 10.1103/PhysRevLett.92.218301} {\bibfield  {journal}
  {\bibinfo  {journal} {Phys. Rev. Lett.}\ }\textbf {\bibinfo {volume} {92}},\
  \bibinfo {pages} {218301} (\bibinfo {year} {2004})}\BibitemShut {NoStop}%
\bibitem [{\citenamefont {Scholes}\ and\ \citenamefont
  {Fleming}(2000)}]{Scholes2000a}%
  \BibitemOpen
  \bibfield  {author} {\bibinfo {author} {\bibfnamefont {G.~D.}\ \bibnamefont
  {Scholes}}\ and\ \bibinfo {author} {\bibfnamefont {G.~R.}\ \bibnamefont
  {Fleming}},\ }\href {\doibase 10.1021/jp993435l} {\bibfield  {journal}
  {\bibinfo  {journal} {J. Phys. Chem. B}\ }\textbf {\bibinfo {volume} {104}},\
  \bibinfo {pages} {1854} (\bibinfo {year} {2000})}\BibitemShut {NoStop}%
\bibitem [{\citenamefont {Sumi}(1974)}]{Sumi1974a}%
  \BibitemOpen
  \bibfield  {author} {\bibinfo {author} {\bibfnamefont {H.}~\bibnamefont
  {Sumi}},\ }\href {\doibase 10.1143/JPSJ.36.770} {\bibfield  {journal}
  {\bibinfo  {journal} {J. Phys. Soc. Jpn.}\ }\textbf {\bibinfo {volume}
  {36}},\ \bibinfo {pages} {770} (\bibinfo {year} {1974})}\BibitemShut
  {NoStop}%
\bibitem [{\citenamefont {Kuhn}(1970)}]{Kuhn1970a}%
  \BibitemOpen
  \bibfield  {author} {\bibinfo {author} {\bibfnamefont {H.}~\bibnamefont
  {Kuhn}},\ }\href {\doibase 10.1063/1.1673749} {\bibfield  {journal} {\bibinfo
   {journal} {J. Chem. Phys.}\ }\textbf {\bibinfo {volume} {53}},\ \bibinfo
  {pages} {101} (\bibinfo {year} {1970})}\BibitemShut {NoStop}%
\bibitem [{\citenamefont {Chance}\ \emph {et~al.}(1975)\citenamefont {Chance},
  \citenamefont {Prock},\ and\ \citenamefont {Silbey}}]{Chance1975a}%
  \BibitemOpen
  \bibfield  {author} {\bibinfo {author} {\bibfnamefont {R.~R.}\ \bibnamefont
  {Chance}}, \bibinfo {author} {\bibfnamefont {A.}~\bibnamefont {Prock}}, \
  and\ \bibinfo {author} {\bibfnamefont {R.}~\bibnamefont {Silbey}},\ }\href
  {\doibase 10.1063/1.430748} {\bibfield  {journal} {\bibinfo  {journal} {J.
  Chem. Phys.}\ }\textbf {\bibinfo {volume} {62}},\ \bibinfo {pages} {2245}
  (\bibinfo {year} {1975})}\BibitemShut {NoStop}%
\bibitem [{\citenamefont {Zimanyi}\ and\ \citenamefont
  {Silbey}(2010)}]{Zimanyi2010a}%
  \BibitemOpen
  \bibfield  {author} {\bibinfo {author} {\bibfnamefont {E.~N.}\ \bibnamefont
  {Zimanyi}}\ and\ \bibinfo {author} {\bibfnamefont {R.~J.}\ \bibnamefont
  {Silbey}},\ }\href {\doibase http://dx.doi.org/10.1063/1.3488136} {\bibfield
  {journal} {\bibinfo  {journal} {J. Chem. Phys.}\ }\textbf {\bibinfo {volume}
  {133}},\ \bibinfo {eid} {144107} (\bibinfo {year} {2010})}\BibitemShut
  {NoStop}%
\bibitem [{\citenamefont {Duque}\ \emph {et~al.}(2015)\citenamefont {Duque},
  \citenamefont {Brumer},\ and\ \citenamefont {Pach\'on}}]{Duque2015a}%
  \BibitemOpen
  \bibfield  {author} {\bibinfo {author} {\bibfnamefont {S.}~\bibnamefont
  {Duque}}, \bibinfo {author} {\bibfnamefont {P.}~\bibnamefont {Brumer}}, \
  and\ \bibinfo {author} {\bibfnamefont {L.~A.}\ \bibnamefont {Pach\'on}},\
  }\href {\doibase 10.1103/PhysRevLett.115.110402} {\bibfield  {journal}
  {\bibinfo  {journal} {Phys. Rev. Lett.}\ }\textbf {\bibinfo {volume} {115}},\
  \bibinfo {pages} {110402} (\bibinfo {year} {2015})}\BibitemShut {NoStop}%
\bibitem [{\citenamefont {Ma}\ \emph {et~al.}(2015)\citenamefont {Ma},
  \citenamefont {Moix},\ and\ \citenamefont {Cao}}]{Ma2015b}%
  \BibitemOpen
  \bibfield  {author} {\bibinfo {author} {\bibfnamefont {J.}~\bibnamefont
  {Ma}}, \bibinfo {author} {\bibfnamefont {J.~M.}\ \bibnamefont {Moix}}, \ and\
  \bibinfo {author} {\bibfnamefont {J.}~\bibnamefont {Cao}},\ }\href {\doibase
  10.1063/1.4908600} {\bibfield  {journal} {\bibinfo  {journal} {J. Chem.
  Phys.}\ }\textbf {\bibinfo {volume} {142}},\ \bibinfo {pages} {094107}
  (\bibinfo {year} {2015})}\BibitemShut {NoStop}%
\bibitem [{\citenamefont {Kampen}(2007)}]{VanKampenBook}%
  \BibitemOpen
  \bibfield  {author} {\bibinfo {author} {\bibfnamefont {N.~V.}\ \bibnamefont
  {Kampen}},\ }\href@noop {} {\emph {\bibinfo {title} {Stochastic Processes in
  Physics and Chemistry}}}\ (\bibinfo  {publisher} {Elsevier},\ \bibinfo {year}
  {2007})\BibitemShut {NoStop}%
\bibitem [{\citenamefont {Chenu}\ and\ \citenamefont {Cao}()}]{CPA_SI2016}%
  \BibitemOpen
  \bibfield  {author} {\bibinfo {author} {\bibfnamefont {A.}~\bibnamefont
  {Chenu}}\ and\ \bibinfo {author} {\bibfnamefont {J.}~\bibnamefont {Cao}},\
  }\href@noop {} {\ }\bibinfo {note} {{S}upplementary Material: link between
  the CPA and the classical approach and further comparison with FCE
  results.}\BibitemShut {Stop}%
\bibitem [{\citenamefont {Yang}(2005)}]{Yang2005a}%
  \BibitemOpen
  \bibfield  {author} {\bibinfo {author} {\bibfnamefont {M.}~\bibnamefont
  {Yang}},\ }\href {\doibase http://dx.doi.org/10.1063/1.2046668} {\bibfield
  {journal} {\bibinfo  {journal} {J. Chem. Phys.}\ }\textbf {\bibinfo {volume}
  {123}},\ \bibinfo {pages} {124705} (\bibinfo {year} {2005})}\BibitemShut
  {NoStop}%
\bibitem [{\citenamefont {Man\v{c}al}\ \emph {et~al.}(2011)\citenamefont
  {Man\v{c}al}, \citenamefont {Balevi\v{c}ius},\ and\ \citenamefont
  {Valkunas}}]{Mancal2011a}%
  \BibitemOpen
  \bibfield  {author} {\bibinfo {author} {\bibfnamefont {T.}~\bibnamefont
  {Man\v{c}al}}, \bibinfo {author} {\bibfnamefont {V.}~\bibnamefont
  {Balevi\v{c}ius}}, \ and\ \bibinfo {author} {\bibfnamefont {L.}~\bibnamefont
  {Valkunas}},\ }\href {\doibase 10.1021/jp108247a} {\bibfield  {journal}
  {\bibinfo  {journal} {J. Phys. Chem. A}\ }\textbf {\bibinfo {volume} {115}},\
  \bibinfo {pages} {3845} (\bibinfo {year} {2011})}\BibitemShut {NoStop}%
\bibitem [{\citenamefont {Cao}\ and\ \citenamefont {Berne}(1993)}]{Cao1993a}%
  \BibitemOpen
  \bibfield  {author} {\bibinfo {author} {\bibfnamefont {J.}~\bibnamefont
  {Cao}}\ and\ \bibinfo {author} {\bibfnamefont {B.~J.}\ \bibnamefont
  {Berne}},\ }\href {\doibase 10.1063/1.465446} {\bibfield  {journal} {\bibinfo
   {journal} {J. Chem. Phys.}\ }\textbf {\bibinfo {volume} {99}},\ \bibinfo
  {pages} {6998} (\bibinfo {year} {1993})}\BibitemShut {NoStop}%
\bibitem [{\citenamefont {Moix}\ \emph {et~al.}(2012)\citenamefont {Moix},
  \citenamefont {Zhao},\ and\ \citenamefont {Cao}}]{Moix2012a}%
  \BibitemOpen
  \bibfield  {author} {\bibinfo {author} {\bibfnamefont {J.~M.}\ \bibnamefont
  {Moix}}, \bibinfo {author} {\bibfnamefont {Y.}~\bibnamefont {Zhao}}, \ and\
  \bibinfo {author} {\bibfnamefont {J.}~\bibnamefont {Cao}},\ }\href {\doibase
  10.1103/PhysRevB.85.115412} {\bibfield  {journal} {\bibinfo  {journal} {Phys.
  Rev. B}\ }\textbf {\bibinfo {volume} {85}},\ \bibinfo {pages} {115412}
  (\bibinfo {year} {2012})}\BibitemShut {NoStop}%
\bibitem [{\citenamefont {Choro\v{s}ajev}\ \emph {et~al.}(2014)\citenamefont
  {Choro\v{s}ajev}, \citenamefont {Gelzinis}, \citenamefont {Valkunas},\ and\
  \citenamefont {Abramavicius}}]{Chorosajev2014a}%
  \BibitemOpen
  \bibfield  {author} {\bibinfo {author} {\bibfnamefont {V.}~\bibnamefont
  {Choro\v{s}ajev}}, \bibinfo {author} {\bibfnamefont {A.}~\bibnamefont
  {Gelzinis}}, \bibinfo {author} {\bibfnamefont {L.}~\bibnamefont {Valkunas}},
  \ and\ \bibinfo {author} {\bibfnamefont {D.}~\bibnamefont {Abramavicius}},\
  }\href {\doibase http://dx.doi.org/10.1063/1.4884275} {\bibfield  {journal}
  {\bibinfo  {journal} {J. Chem. Phys.}\ }\textbf {\bibinfo {volume} {140}},\
  \bibinfo {eid} {244108} (\bibinfo {year} {2014})}\BibitemShut {NoStop}%
\bibitem [{\citenamefont {Cheng}\ and\ \citenamefont
  {Silbey}(2006)}]{Cheng2006a}%
  \BibitemOpen
  \bibfield  {author} {\bibinfo {author} {\bibfnamefont {Y.~C.}\ \bibnamefont
  {Cheng}}\ and\ \bibinfo {author} {\bibfnamefont {R.~J.}\ \bibnamefont
  {Silbey}},\ }\href {\doibase 10.1103/PhysRevLett.96.028103} {\bibfield
  {journal} {\bibinfo  {journal} {Phys. Rev. Lett.}\ }\textbf {\bibinfo
  {volume} {96}},\ \bibinfo {pages} {028103} (\bibinfo {year}
  {2006})}\BibitemShut {NoStop}%
\bibitem [{\citenamefont {Kim}\ \emph {et~al.}(2010)\citenamefont {Kim},
  \citenamefont {Chang}, \citenamefont {Korenblit}, \citenamefont {Islam},
  \citenamefont {Edwards}, \citenamefont {Freericks}, \citenamefont {Lin},
  \citenamefont {Duan},\ and\ \citenamefont {Monroe}}]{Kim2010b}%
  \BibitemOpen
  \bibfield  {author} {\bibinfo {author} {\bibfnamefont {K.}~\bibnamefont
  {Kim}}, \bibinfo {author} {\bibfnamefont {M.-S.}\ \bibnamefont {Chang}},
  \bibinfo {author} {\bibfnamefont {S.}~\bibnamefont {Korenblit}}, \bibinfo
  {author} {\bibfnamefont {R.}~\bibnamefont {Islam}}, \bibinfo {author}
  {\bibfnamefont {E.~E.}\ \bibnamefont {Edwards}}, \bibinfo {author}
  {\bibfnamefont {J.~K.}\ \bibnamefont {Freericks}}, \bibinfo {author}
  {\bibfnamefont {G.-D.}\ \bibnamefont {Lin}}, \bibinfo {author} {\bibfnamefont
  {L.-M.}\ \bibnamefont {Duan}}, \ and\ \bibinfo {author} {\bibfnamefont
  {C.}~\bibnamefont {Monroe}},\ }\href@noop {} {\bibfield  {journal} {\bibinfo
  {journal} {Nature}\ }\textbf {\bibinfo {volume} {465}},\ \bibinfo {pages}
  {590} (\bibinfo {year} {2010})}\BibitemShut {NoStop}%
\bibitem [{\citenamefont {Wu}\ \emph {et~al.}(2013)\citenamefont {Wu},
  \citenamefont {Silbey},\ and\ \citenamefont {Cao}}]{Wu2013b}%
  \BibitemOpen
  \bibfield  {author} {\bibinfo {author} {\bibfnamefont {J.}~\bibnamefont
  {Wu}}, \bibinfo {author} {\bibfnamefont {R.~J.}\ \bibnamefont {Silbey}}, \
  and\ \bibinfo {author} {\bibfnamefont {J.}~\bibnamefont {Cao}},\ }\href
  {\doibase 10.1103/PhysRevLett.110.200402} {\bibfield  {journal} {\bibinfo
  {journal} {Phys. Rev. Lett.}\ }\textbf {\bibinfo {volume} {110}},\ \bibinfo
  {pages} {200402} (\bibinfo {year} {2013})}\BibitemShut {NoStop}%
\bibitem [{\citenamefont {Friesner}\ and\ \citenamefont
  {Silbey}(1982)}]{Friesner1982a}%
  \BibitemOpen
  \bibfield  {author} {\bibinfo {author} {\bibfnamefont {R.}~\bibnamefont
  {Friesner}}\ and\ \bibinfo {author} {\bibfnamefont {R.~J.}\ \bibnamefont
  {Silbey}},\ }\href {\doibase 10.1016/0009-2614(82)83674-9} {\bibfield
  {journal} {\bibinfo  {journal} {Chem. Phys. Lett.}\ }\textbf {\bibinfo
  {volume} {93}},\ \bibinfo {pages} {107} (\bibinfo {year} {1982})}\BibitemShut
  {NoStop}%
\bibitem [{\citenamefont {Lagos}\ and\ \citenamefont
  {Friesner}(1984)}]{Lagos1984a}%
  \BibitemOpen
  \bibfield  {author} {\bibinfo {author} {\bibfnamefont {R.~E.}\ \bibnamefont
  {Lagos}}\ and\ \bibinfo {author} {\bibfnamefont {R.~A.}\ \bibnamefont
  {Friesner}},\ }\href {\doibase 10.1103/PhysRevB.29.3045} {\bibfield
  {journal} {\bibinfo  {journal} {Phys. Rev. B}\ }\textbf {\bibinfo {volume}
  {29}},\ \bibinfo {pages} {3045} (\bibinfo {year} {1984})}\BibitemShut
  {NoStop}%
\end{thebibliography}
% --- BIBLIO ---

% ------------------

\appendix

\clearpage

%%%%%%%%%% Prefix a "S" to all equations, figures, tables and reset the counter %%%%%%%%%%
\setcounter{equation}{0}
\setcounter{figure}{0}
\setcounter{table}{0}
\makeatletter
\renewcommand{\theequation}{S\arabic{equation}}
\renewcommand{\thefigure}{S\arabic{figure}}
%%%%%%%%%% Prefix a "S" to all equations, figures, tables and reset the counter %%%%%%%%%%

\section{Appendix}

\emph{Derivation of the CPA absorption tensor.---}
We give here the details that lead to the total Green's function (11) used to obtain the absorption tensor (12). 
We start from the definition of the total Green's function in the time domain (7a) 
and write the last exponential using the integral representation 
\begin{equation}
e^{- i \frac{H}{\hbar} t} = \int\limits_{-\infty}^{\infty} \frac{d\omega}{-i 2\pi} \frac{e^{- i (\omega + i 0_+)t}}{\omega  - \frac{(H_0 + V)}{\hbar} + i 0_+}.
\end{equation}
Because the solution for the non-interacting monomers is known, we use the fact that the Hamiltonian splits into $H = H_0 + V$ and obtain the response function from iteration of the identity:
\begin{equation}
\frac{1}{\omega - H} = \frac{1}{\omega - H_0} + \frac{1}{\omega- H} V \frac{1}{\omega- H_0}.
\end{equation}
Using the convolution theorem and transforming in the Fourier domain, the total Green's function  becomes 
\begin{eqnarray}
\mathcal{G}_{nn'}(\omega) &= &\mathcal{G}^0_{nn'}(\omega)  +   \mathcal{G}^0_{nn}(\omega) V_{nn'}  \mathcal{G}^0_{n'n'}(\omega) \\
&&+\sum_{m }\mathcal{G}^0_{nn}(\omega)  V_{nm}  \mathcal{G}^0_{mm}(\omega) V_{mn'}  \mathcal{G}^0_{n'n'}(\omega) + \dots. \nonumber
\end{eqnarray}
The first term on the l.h.s. is only non-zero for $\delta_{nn'}$. Hence we see that the total Green's function only depends on the monomeric function $\mathcal{G}^0$ in-between the interactions in that all functions are diagonal in the system basis.  Taking the thermal average over the phonon bath with an initial ground state and using the approximation (10) yields to the matrix representation (8), from which we can obtain the total absorption tensor (12). 
\\
\\
\emph{Link to the classical MCFT.---} In a recent study [38], the MCFT was derived from classical electrodynamics. Using the picture of classical oscillating dipoles, the  MCFT rate (16) was recast into the overlap of an equivalent emission and absorption spectra, defining
\begin{subequations}
\begin{align}
E^D_{mm'}(\omega) &=  (p^D_m(\omega))^* \, (p^D_{m'}(\omega)) \label{eq:E_CED} \\
I^A_{nn'}(\omega) &= \omega \: {\rm Im}[\boldsymbol{\chi}_A^{-1}(\omega)/\epsilon_0 - \Phi^A]^{-1}_{nn'}, \label{eq:I_CED}
\end{align}
\end{subequations}
where $p^D_m(\omega)$ denotes the polarization of the $m$-th donor molecule. $\boldsymbol{\chi}(\omega)$ is the polarizability tensor and relates the polarization to the total electric field (external and induced) as $p_n(\omega) = \epsilon_0 \chi_n(\omega) E(\br_n,\omega)$. The dipoles are coupled through dipole-dipole coupling, denoted by the $\Phi$ matrix, such that $E(\br_n, \omega) = E_{\rm ext}(\br_n, \omega) + \sum_{nn'} \Phi_{n n'} p_{n'}(\omega)$. 
 Considering that only the donor complex is excited by the external field, the classically-defined absorption spectra (\ref{eq:I_CED}) from [38] takes the form 
\begin{equation}
\mI_{\rm CED}(\omega) = \omega {\rm Im} \left [ \frac{\epsilon_0 \boldsymbol{\chi}_A (\omega)}{\mathbbm{1} - \boldsymbol{\Phi}^A \epsilon_0 \boldsymbol{\chi}_A (\omega) } \right]
\end{equation}
This matches the derived absorption tensor used in (17), up to a normalizing factor and a factor $\omega$ arising for the different definitions between susceptibility and absorption coefficient, taking $\langle \mathbfcal{G}^0_A(\omega) \rangle \rightarrow {\epsilon_0} \boldsymbol{\chi}_A (\omega)$ and $\mV^A \rightarrow \boldsymbol{\Phi}^A$, i.e. when we restrict the electronic coupling in our model Hamiltonian (6) to dipole-dipole interaction. 

\begin{figure}
\includegraphics[width=\columnwidth]{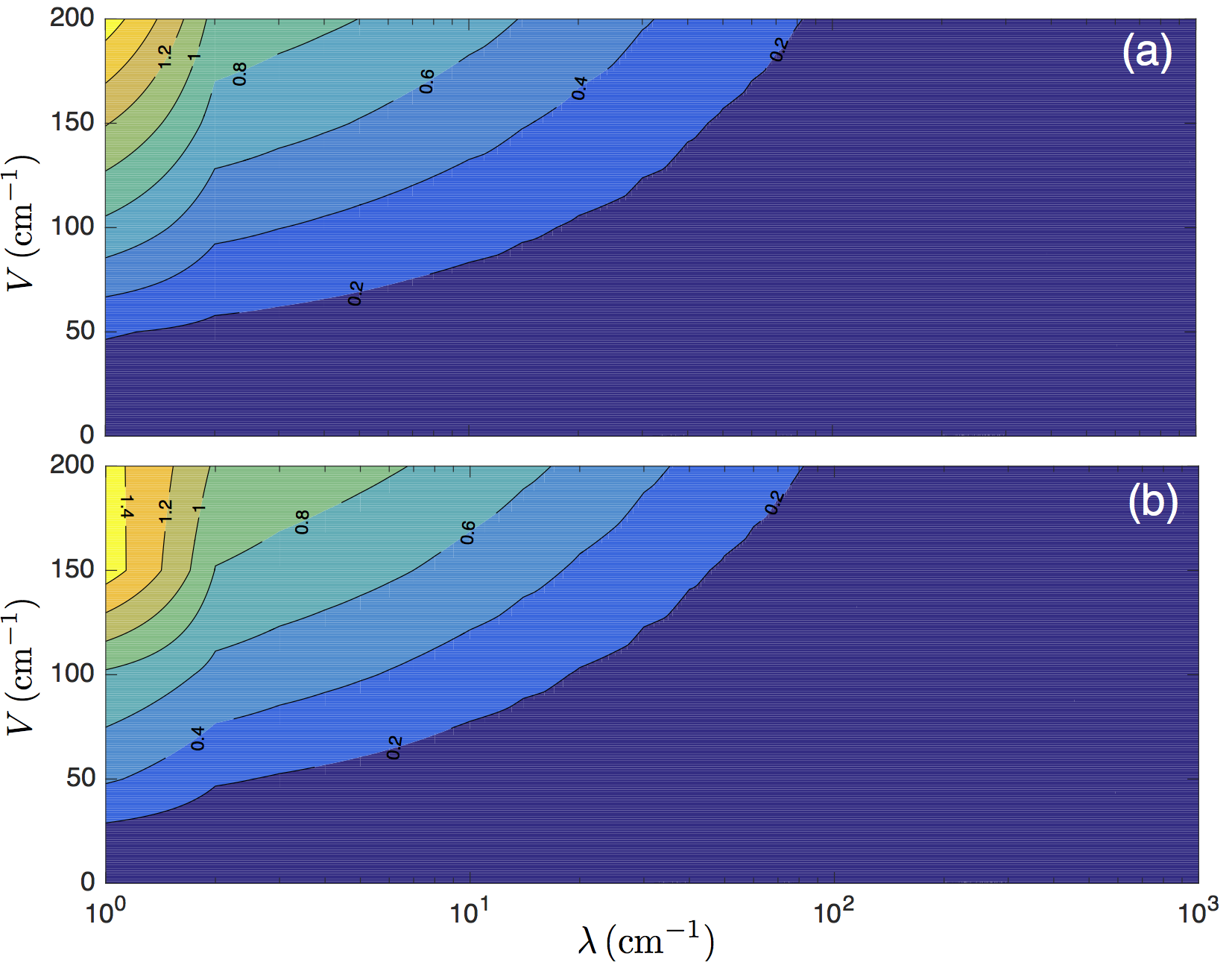}
\caption{Relative difference  $\epsilon_{\rm CPA}$  (\%) of the CPA method compared to the FCE model for the absorption spectra calculated from the tensor (12) as a function of the reorganization energy $\lambda$ and the intra-aggregate coupling $V_{nn'}=V$ showing very good agreement. (a) $\Delta E_{12}$ = 100 cm$^{-1}$ and (b) $\Delta E_{12}$ = 20 cm$^{-1}$. \label{fig:errAbs}}
\bigskip
\includegraphics[width=\columnwidth]{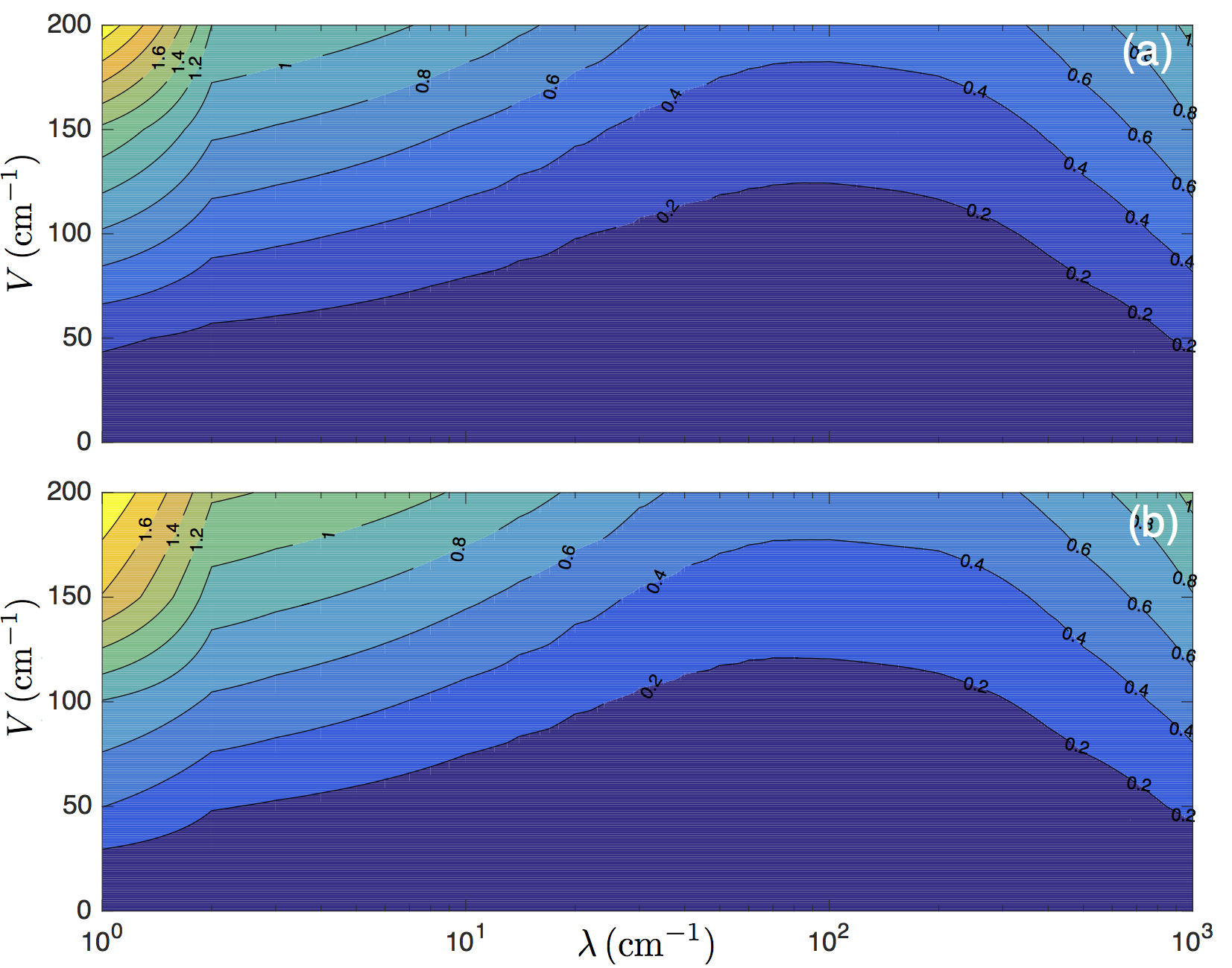}
\caption{Relative difference  $\epsilon_{\rm CPA_{DB}}$   (\%) of the CPA$_{\rm DB}$ method compared to the FCE model for the emission spectra calculated from the tensor (14) as a function of the reorganization energy $\lambda$ and the intra-aggregate coupling $V_{nn'}=V$ showing very good agreement.  (a) $\Delta E_{12}$ = 100 cm$^{-1}$ and (b) $\Delta E_{12}$ = 20 cm$^{-1}$. \label{fig:errEmi}}
\end{figure}

\begin{figure}
\includegraphics[width=\columnwidth]{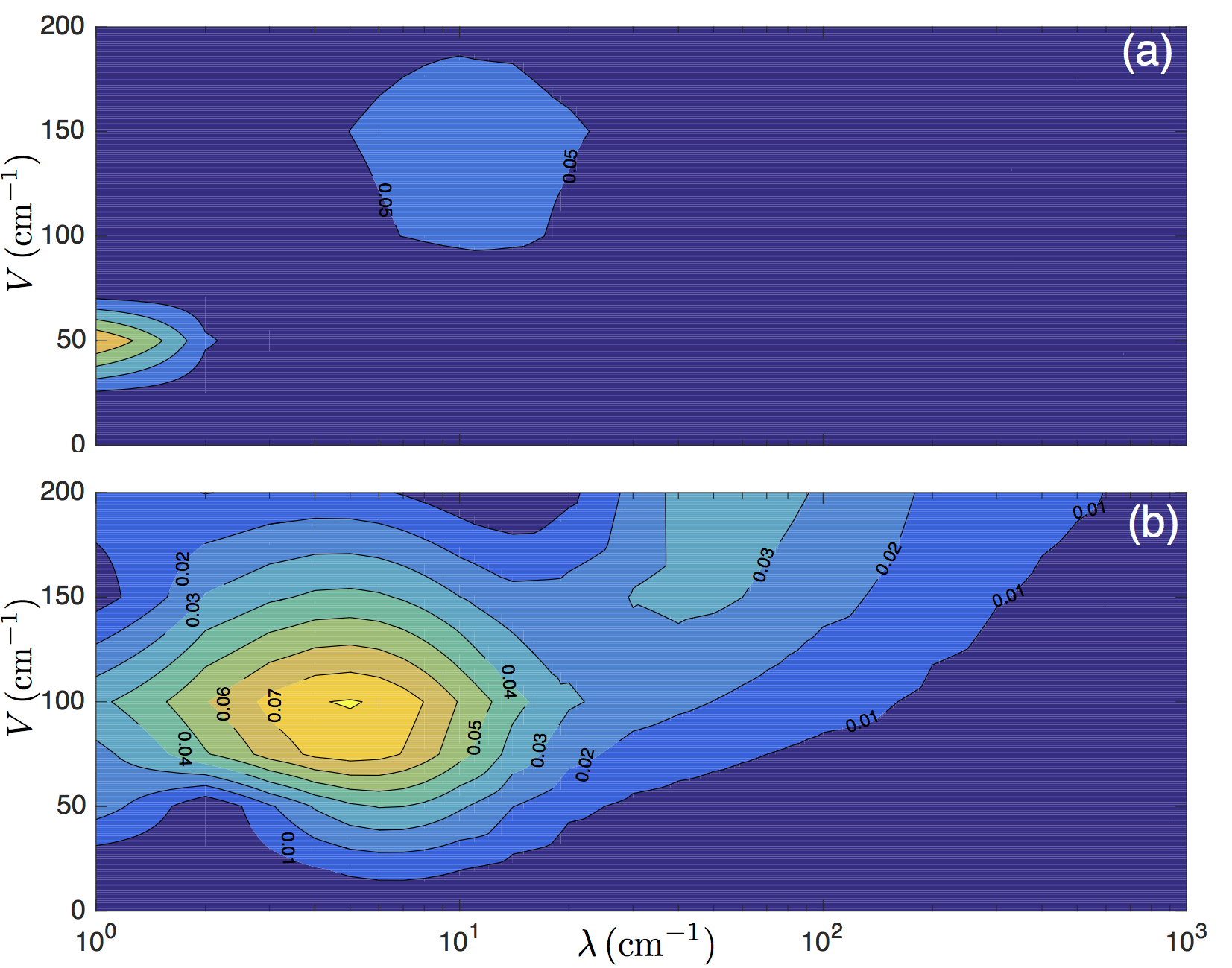}
\caption{ Relative difference   $\epsilon_{\rm CPA_{DB}}$  (\%) of the CPA$_{\rm DB}$ method compared to the FCE model for the transfer rate (17) as a function of the reorganization energy $\lambda$ and the intra-aggregate coupling $V_{nn'}=V$ showing very good agreement. (a) $\Delta E_{12}$ = 100 cm$^{-1}$ and (b) $\Delta E_{12}$ = 20 cm$^{-1}$.  \label{fig:errRate_map}}
\end{figure}

Emission, in turn, is a purely quantum phenomena. We look here at the classically defined emission spectra to see how it compares with the quantum definition derived in this paper. 
First-order response theory allows to obtain the classical polarization from the total field. Solving the system of linear equations yields
\begin{equation}
\mathbf{p} (\omega) = \frac{\epsilon_0  \boldsymbol{\chi}(\omega)}{ \mathbbm{1} - \Phi \epsilon_0 \boldsymbol{\chi}(\omega)} \cdot \bE_{\rm ext}(\omega).
\end{equation}
The classically defined emission spectra (\ref{eq:E_CED}) then depends on the polarization of the acceptor through the susceptibility tensor. This is well understood by the Purcell effect, according to which the polarization of a dipole in a medium depends on scattered electrical field. However, it is different from the model developed here, where the donor emission spectrum can be defined independently of the acceptor state, and where the donor-acceptor coupling only enters in the rate equation. 
\\
\\
\emph{Further results.---} We present here further comparison of the CPA approximation with the FCE [22] for the absorption (Fig. \ref{fig:errAbs}) and emission spectra (Fig. \ref{fig:errEmi}), and for the rate (Fig. \ref{fig:errRate_map}), respectively given by Eqs. (12),  (14)  and  (17). 
We define the `relative difference' (in \%) between the experimental spectra obtained from the method `X' and the exact spectra as obtained reliably from the FCE as, for absorption,  
\begin{equation}
\epsilon_X = \frac{\int d\omega \: | I_{\rm FCE}(\omega) - I_{\rm X} (\omega)| }{\int d\omega \: I_{\rm FCE}(\omega)},
\end{equation}
and equivalently for the emission with $I(\omega) \rightarrow E(\omega)$. 
The electronic energy splitting is (a) $\Delta E_{12}$ = 100 cm$^{-1}$ and (b) $\Delta E_{12}$ = 20 cm$^{-1}$ in all figures below. Very good agreement is found for all the tested regimes, with less that 2\% error between the FCE and CPA$_{\rm DB}$ results.

\clearpage

\end{document}